\documentclass[preprint,3p,12p]{elsarticle}

\usepackage{graphicx}
\usepackage{atlasphysics}
\usepackage{amsmath,amssymb,amstext}
\usepackage{multirow}
\usepackage{wasysym}  
\usepackage{color}	  
\usepackage[colorlinks,breaklinks,pdfauthor={The ATLAS Collaboration}]{hyperref}  
\hypersetup{linkcolor=blue,citecolor=blue,filecolor=black,urlcolor=blue}

\def\betastar{\ensuremath{\beta^{\star}}}
\def\sigmatot{\ensuremath{\sigma_{\mathrm{tot}}}}

\biboptions{comma,square,sort&compress}

\journal{Nuclear Physics B}

\usepackage{preprintcover} 

\PreprintCoverPaperTitle{Measurement of the total cross section from elastic 
scattering in $pp$ collisions at $\rts=7\TeV$ with the ATLAS detector}  
\PreprintIdNumber{CERN-PH-EP-2014-177}
\PreprintCoverAbstract{A measurement of the total $pp$ cross section at the LHC at $\rts=7\TeV$ is presented. 
In a special run with high-\betastar\, beam optics, an integrated luminosity of 
$80$ $\mu$b$^{-1}$ was accumulated in order to measure the differential elastic cross section as a function of 
the Mandelstam momentum transfer variable $t$. The measurement is performed with the ALFA sub-detector of ATLAS. 
Using a fit to the differential elastic cross section in the $|t|$ range from $0.01~\GeV^2$ to $0.1~\GeV^2$ to 
extrapolate to $|t|\rightarrow 0$, the total cross section, $\sigmatot(pp\rightarrow X)$,  
is measured via the optical theorem to be:
\begin{equation*}
\sigmatot(pp\rightarrow X) =  95.35 \; \pm 0.38 \; ({\mbox{stat.}}) \pm 1.25 \; 
({\mbox{exp.}}) \pm 0.37 \; (\mbox{extr.})  \; \mbox{mb} \; \; , 
\end{equation*}
where the first error is statistical, the second accounts for all experimental systematic uncertainties and the last 
is related to uncertainties in the extrapolation to $|t|\rightarrow 0$. 
In addition, the slope of the elastic cross section at small $|t|$ is determined to be $B = 19.73 \pm 0.14 \; ({\mbox{stat.}}) \pm 0.26  \; ({\mbox{syst.}}) \; \mbox{\GeV}^{-2}$.
}
\PreprintJournalName{Nuclear Physics B}

\renewcommand*{\today}{November 3, 2014} 

\begin{document}

\begin{frontmatter}
\title{Measurement of the total cross section from elastic 
scattering in $pp$ collisions at $\rts=7\TeV$ with the ATLAS detector}
\author{The ATLAS Collaboration}
\ead{atlas.publications@cern.ch}
\address{CERN, 1211 Geneva 23, Switzerland}

\begin{abstract}
A measurement of the total $pp$ cross section at the LHC at $\rts=7\TeV$ is presented. 
In a special run with high-\betastar\, beam optics, an integrated luminosity of 
$80$ $\mu$b$^{-1}$ was accumulated in order to measure the differential elastic cross section as a function of 
the Mandelstam momentum transfer variable $t$. The measurement is performed with the ALFA sub-detector of ATLAS. 
Using a fit to the differential elastic cross section in the $|t|$ range from $0.01~\GeV^2$ to $0.1~\GeV^2$ to 
extrapolate to $|t|\rightarrow 0$, the total cross section, $\sigmatot(pp\rightarrow X)$,  
is measured via the optical theorem to be:
\begin{equation*}
\sigmatot(pp\rightarrow X) =  \mbox{95.35} \; \pm 0.38 \; ({\mbox{stat.}}) \pm 1.25 \; 
({\mbox{exp.}}) \pm 0.37 \; (\mbox{extr.})  \; \mbox{mb} \; \; , 
\end{equation*}
where the first error is statistical, the second accounts for all experimental systematic uncertainties and the last 
is related to uncertainties in the extrapolation to $|t|\rightarrow 0$. 
In addition, the slope of the elastic cross section at small $|t|$ is determined to be $B = 19.73 \pm 0.14 \; ({\mbox{stat.}}) \pm 0.26  \; ({\mbox{syst.}}) \; \mbox{\GeV}^{-2}$.    
\end{abstract}


\end{frontmatter}


\section{Introduction}
\label{sec:intro}
The total hadronic cross section is a fundamental parameter of strong interactions, setting the scale 
of the size of the interaction region  at a given energy. 
A calculation of the total hadronic cross section from first principles, based upon quantum chromodynamics (QCD), 
is currently not possible. 
Large distances are involved in the collision process and thus perturbation theory is not applicable. 
Even though the total cross section cannot be directly 
calculated, it can be estimated or bounded by a number of fundamental relations in high-energy scattering theory which 
are  model independent. The  Froissart--Martin bound~\cite{froissart,martin1}, which states that the 
total cross section cannot grow asymptotically 
faster than $\ln^2s$, $\sqrt{s}$ being the centre-of-mass energy, is based upon principles of axiomatic field theory. 
The optical theorem, which relates 
the imaginary part of the forward elastic-scattering amplitude to the total cross section, is a general theorem in quantum scattering theory. 
Dispersion relations, which connect the real part of the elastic-scattering amplitude to an integral of the total cross section 
over energy, are based upon the analyticity and crossing symmetry of the scattering amplitude. 
All of these relations lead to testable constraints on the total cross section. 
\par
The rise of the $pp$ cross section with energy was first observed at the ISR~\cite{Pis1,Amaldi2}. 
The fact that the hadronic cross section continues to rise has been confirmed in every new energy regime made accessible 
by a new $pp$ or $\antibar{p}$ collider (Sp$\mathrm{\bar{p}}$S, Tevatron and LHC)~\cite{UA4,UA42,E710A,E811,CDF,TOTEM_first,TOTEM_second}. 
However, the ``asymptotic" energy dependence is yet to be determined. 
A still open question is whether the cross section indeed rises proportionally to $\ln^ 2s$ 
in order to saturate the Froissart--Martin bound or whether the rise has e.g. a $\ln s$ dependence.

\par
Traditionally, the total cross section at hadron colliders has been measured via elastic scattering using the optical theorem. 
This paper presents a measurement by the ATLAS experiment~\cite{atlas1}  at the LHC in $pp$ collisions at $\sqrt{s}=7 \TeV$  using this approach.
The optical theorem states:
\begin{equation}\label{eq:OpticalTheorem}
\sigmatot \propto \mbox{Im} \, [ f_{\mathrm{el}}\left(t \rightarrow 0\right)] \, 
\end{equation}
where $f_{\mathrm{el}}(t \rightarrow 0)$ is the elastic-scattering amplitude extrapolated to the forward direction, 
i.e.\ at  $|t|\rightarrow 0$, 
$t$ being the four-momentum transfer.  
Thus, a measurement of elastic scattering in the very forward direction  gives information on the total cross section.   
An independent measurement of the luminosity is required. In this analysis, the luminosity is determined from LHC beam parameters using 
van der Meer scans~\cite{svdm}.
Once the luminosity is known, the elastic cross section can be normalized. An extrapolation of the differential cross section 
to $|t| \rightarrow 0$ gives the total cross section through the formula:
\begin{equation}\label{eq:totxs}
\sigmatot^{2} = \frac{16\pi(\hbar c)^2}{1+\rho^2} \left. \frac{\mathrm{d}\sigma_{\mathrm{el}}}{\mathrm{d}t}\right|_{t \rightarrow 0} \; ,  
\end{equation}
where $\rho$  represents a small correction arising from the ratio of the real to 
imaginary part of the elastic-scattering amplitude in the forward direction and is taken from theory. 
In order to minimize the model dependence in the extrapolation to $|t|\rightarrow 0$, 
the elastic cross section has to be measured down to as small $|t|$ values as possible. 
Here, a fit in the range $0.01 \GeV^2 < -t < 0.1 \GeV^2$ is used to extract the 
total cross section, while the differential cross section is measured in the range $0.0025 \GeV^2 < -t < 0.38 \GeV^2$. 
The determination of the total cross section also implies a measurement of the nuclear slope parameter $B$, 
which  describes  the exponential $t$-dependence of the nuclear amplitude at small $t$-values. In a simple geometrical model of 
elastic scattering, $B$ is related to the size of the proton and thus its energy dependence 
is strongly correlated with that of the total cross section. 
\par
The measurements of the total cross section and elastic scattering reported here are used to determine 
the inelastic cross section, as the difference between these two quantities. 
This measurement of the inelastic cross section is compared with a previous measurement by the ATLAS experiment using a 
complementary method based upon data from a minimum-bias trigger~\cite{ATLAS_inel}. The ratio of the elastic to total cross section is also derived. 
In the black-disc limit, the limit in which the proton is completely opaque, this quantity goes at asymptotic energies to 0.5 and thus the measurement is directly 
sensitive to the hadron opacity. 
The quantities measured and reported here have also been measured at the LHC by the 
TOTEM experiment~\cite{totem1,TOTEM_lumindep}. 

\par
This paper is organized as follows. 
The experimental setup including a brief description of the ALFA sub-detector 
is given in Section~\ref{sec:exp}, followed by a short description of the measurement method in Section~\ref{sec:meth}.
Section~\ref{sec:mc_th} summarizes theoretical predictions and Monte Carlo simulations. 
The data taking and trigger conditions are outlined in Section~\ref{sec:data_taking}, followed by 
a description of the track reconstruction and alignment procedures in Section~\ref{sec:tracking}. 
The data analysis consisting of event selection, background determination and reconstruction 
efficiency is explained in Section~\ref{sec:data_analysis}. Section~\ref{sec:unfolding_acceptance} describes the acceptance and unfolding corrections. 
The determination of the beam optics is summarized 
in Section~\ref{sec:optics} and of the luminosity in Section~\ref{sec:luminosity}. 
Results for the differential elastic cross section 
are reported in Section~\ref{sec:tspect} and the extraction of the total cross section in Section~\ref{sec:sigma_tot}.  
The results are discussed in Section~\ref{sec:discussion}  with a summary in Section~\ref{sec:conclusion}.

\section{Experimental setup}
\label{sec:exp}
ATLAS is a multi-purpose detector designed to study elementary processes in 
proton--proton interactions at the$\TeV$ energy scale. It consists of an inner 
tracking system, calorimeters and a muon spectrometer surrounding the interaction 
point of the colliding beams. 
The tracking system covers the pseudorapidity range $|\eta| < 2.5$ and the calorimetric 
measurements range to $|\eta| = 4.9$.\footnote{ATLAS uses a right-handed coordinate system 
  with its origin at the nominal 
  interaction point in the centre of the detector and the $z$-axis along the beam pipe. 
  The $x$-axis points from the interaction point to the centre of the LHC ring and the 
  $y$-axis points upwards.
  The pseudorapidity $\eta$ is defined in terms of the polar
  angle $\theta$ as $\eta = -\text{ln tan}(\theta/2)$.}
To improve the coverage in the forward 
direction three smaller detectors with specialized tasks are installed at large 
distance from the interaction point. The most forward detector, ALFA, is sensitive to 
particles in the range $|\eta| > 8.5$, while the two others have acceptance 
windows at $|\eta| \approx 5.8$ (LUCID) and $|\eta| \approx 8.2$ (ZDC). 
A detailed description of the ATLAS detector can be found in Ref.~\cite{atlas1}.

The ALFA detector (Absolute Luminosity For ATLAS) 
is designed to measure small-angle proton scattering. 
Two tracking stations are placed on each side of the central ATLAS 
detector at distances of 238 m and 241 m from the interaction point. 
The tracking detectors are housed in so-called Roman Pots (RPs) which can be moved close 
to the circulating proton beams. Combined with special beam optics, as introduced in 
Section~\ref{sec:meth}, 
this allows the detection of protons at scattering angles down to 
10 $\mu$rad. 

\begin{figure}[h]
  \centering
  \includegraphics[width=80mm]{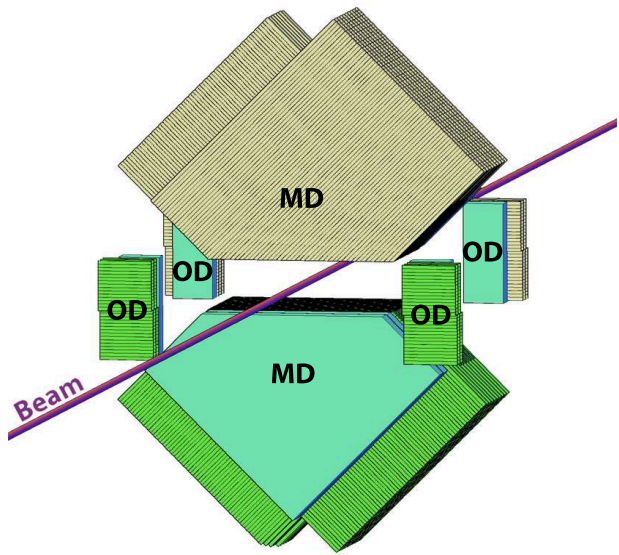} 
  \caption{A schematic view of a pair of ALFA tracking detectors in the upper and lower RPs.
           Although not shown, the ODs on either side of each MD are mechanically 
           attached to them. 
           The orientation of the scintillating fibres is indicated by dashed lines. 
           The plain objects visible in front of the lower MD and ODs are the 
           trigger counters. For upper MD and the lower ODs they are hidden at the opposite 
           side of the fibre structures.}   
  \label{fig:ALFA_front_view}
\end{figure}

Each station carries an upper and lower RP connected by flexible bellows 
to the primary LHC vacuum. The RPs are made of stainless steel with thin windows 
of 0.2 mm and 0.5 mm thickness at the bottom and front sides to reduce the interactions 
of traversing protons.
Elastically scattered protons are detected in the main detectors (MDs) while dedicated overlap 
detectors (ODs) measure the distance between upper and lower MDs. 
The arrangement of the upper and lower MDs and ODs with respect to the beam 
is illustrated in Fig.~\ref{fig:ALFA_front_view}. 

Each MD consists of 2 times 10 layers of 64 
square scintillating fibres with 0.5 mm side length 
glued on titanium plates. The fibres on the front and back sides of each titanium plate 
are orthogonally arranged at angles of $\pm$45$^\circ$ with respect to the $y$-axis.
The projections perpendicular to the fibre axes define the   
$u$ and $v$ coordinates which are used in the track reconstruction described 
in Section~\ref{sec:track_reco}. 
To minimize optical cross-talk, each fibre is coated with a thin aluminium film. The 
individual fibre layers are staggered by multiples of 1/10 of the fibre size 
to improve the position resolution. 
The theoretical resolution of 14.4 $\mu$m per $u$ or $v$ coordinate is degraded  
due to imperfect staggering, cross-talk, noise and inefficient fibre channels.
To reduce the impact of imperfect staggering on
the detector resolution, all fibre positions were measured by microscope.
In a test beam~\cite{testbeam1,testbeam2} with 120$\GeV$ hadrons, the position resolution
was measured to be between 30 $\mu$m and 35 $\mu$m. The efficiency to detect a traversing 
proton in a single fibre layer is typically 93$\%$, with layer-to-layer variations of 
about 1$\%$. 
The overlap detectors consist of three layers of 30 scintillating fibres per layer measuring 
the vertical coordinate of traversing beam-halo particles or shower fragments.\footnote{Halo 
particles originate from beam 
particles which left the bunch structure of the beam but still circulate in the beam pipe.}
Two independent ODs are attached at each side of both MDs,
as sketched in Fig.~\ref{fig:ALFA_front_view}.
The alignment of the ODs with respect to the coordinate system of the MDs was performed 
by test-beam measurements using a silicon pixel telescope. 
A staggering by 1/3 of the fibre size results in a single-track 
resolution of about 50 $\mu$m. 
The signals from both types of tracking detectors are amplified by 
64-channel 
multi-anode photomultipliers (MAPMTs). 
The scintillating fibres are directly coupled to the MAPMT photocathode.  
Altogether, 23 MAPMTs are used to read out each MD and its two adjacent ODs.

Both tracking detectors are completed by trigger counters 
which consist of 3 mm thick scintillator plates covering the active areas of MDs and ODs. 
Each MD is equipped with two trigger counters and their signals are used in coincidence to
reduce noise contributions. The ODs are 
covered by single trigger counters and each signal is recorded. Clear-fibre bundles are 
used to guide all scintillation signals from the trigger counters to 
single-channel photomultipliers. 

Before data taking, precision motors move the RPs vertically in 5 $\mu$m steps towards 
the beam. The position measurement is realized by inductive 
displacement sensors (LVDT) 
calibrated by a laser survey in the LHC tunnel. The internal precision of these 
sensors is 10 $\mu$m. In addition, the motor steps are used to cross-check the 
LVDT values. 

The compact front-end electronics is assembled in a three-layer structure attached to the 
back side of each MAPMT. The three layers comprise a high-voltage 
divider board, a passive board for signal routing and an active board for signal amplification, 
discrimination and buffering using the MAROC chip~\cite{MAROC1,MAROC2}. 
The buffers of all 23 MAPMT readout chips of a complete detector are serially transmitted 
by five kapton cables to the mother-board.
All digital signals are transmitted via a fibre optical link to the central ATLAS 
data acquisition system. The analogue trigger signals are sent by fast air-core cables to 
the central trigger processor.   

\begin{figure}[hb!]
  \centering
  \includegraphics[width=\textwidth]{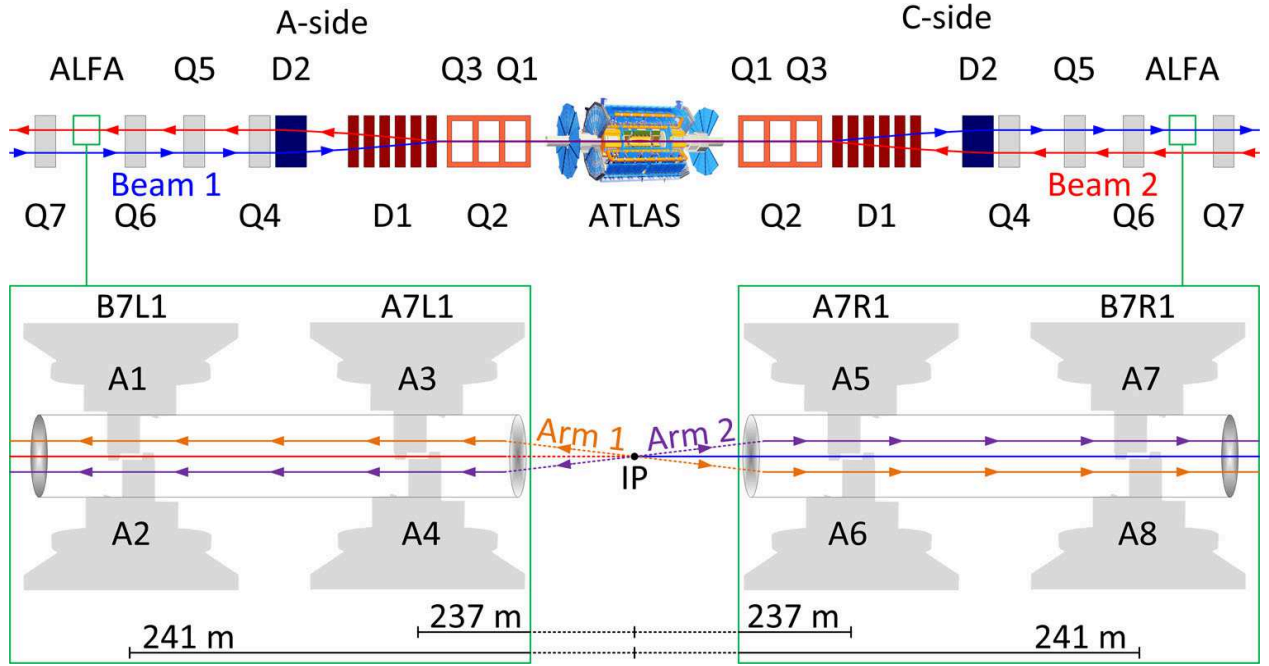} 
  \caption{A sketch of the experimental set-up, not to scale, showing the
   positions of the ALFA Roman Pot stations in the outgoing LHC beams, and the quadrupole
   (Q1--Q6) and dipole (D1--D2) magnets situated between the interaction point and ALFA.
   The ALFA detectors are numbered A1--A8, and are combined into inner stations A7R1 and A7L1, 
   which are closer to the interaction point, and outer stations B7R1 and B7L1. 
   The arrows indicate in the top panel the beam directions and in the bottom panel the scattered 
   proton directions.}
  \label{fig:alfa_layout}
\end{figure}

The station and detector naming scheme is depicted in Fig.~\ref{fig:alfa_layout}.
The stations A7R1 and B7R1 are positioned at $z$ = $-$237.4 m and $z$ = $-$241.5 m respectively 
in the outgoing beam 1 (C side), while the stations A7L1 and B7L1  are situated 
symmetrically in the outgoing beam 2 (A side). 
The detectors A1--A8 are inserted in 
increasing order in stations B7L1, A7L1, A7R1 and B7R1 with even-numbered detectors in the 
lower RPs. Two spectrometer arms for elastic-scattering event topologies are defined by the following 
detector series: arm 1 comprising detectors A1, A3, A6, A8, and arm 2 comprising 
detectors A2, A4, A5, A7.
The sequence of quadrupoles between the interaction point and ALFA is also shown in 
Fig.~\ref{fig:alfa_layout}. Among them, the inner triplet Q1--Q3 is most important for the 
high-$\betastar$ beam optics necessary for this measurement.

\section{Measurement method}
\label{sec:meth}
The data were recorded with special beam optics characterized by 
 a $\betastar$ of 90 m~\cite{Note90m,HBLumiday11} at the interaction point resulting in a small divergence and 
providing parallel-to-point focusing in the vertical plane.\footnote{The $\beta$-function determines the variation of 
the beam envelope around the ring and depends on the focusing properties of the magnetic lattice.} 
In parallel-to-point beam optics  
the betatron oscillation has a phase advance $\Psi$ of $90^\circ$ between the interaction point 
and the RPs, such that all particles scattered 
at the same angle are focused at the same position at the detector, independent of their production vertex position. 
This focusing is only achieved in the vertical plane. 

The beam optics parameters are needed for the reconstruction of the scattering angle 
$\theta^\star$ at the interaction point. 
The four-momentum transfer $t$ is calculated from $\theta^\star$; 
in elastic scattering at high energies this is given by:
\begin{equation}
\label{eq:t-basic}
 -t = \left(\theta^\star \times p \right)^2 \,\,,
\end{equation}
where $p$ is the nominal beam momentum of the LHC of $3.5 \TeV$ and 
$\theta^\star$ is measured from the proton tra\-jectories in ALFA.  
A formalism based on transport matrices 
allows positions and angles of particles at two different points of the magnetic lattice to be related.

The trajectory ($w(z)$, $\theta_w(z)$), where $w\in\{x,y\}$ is the transverse position with 
respect to the nominal orbit at a distance $z$ from 
the interaction point and $\theta_w$ is the angle between $w$ and $z$, is 
given by the transport matrix {\bf M} and the coordinates at 
the interaction point ($w^\star$, $\theta_w^\star$):

\begin{equation}\label{eq:transport}
\left(\begin{array}{c}
w(z) \\
\theta_w(z) \\
\end{array}\right) =
{\bf M} 
\left(\begin{array}{c}
w^\star \\
\theta_w^\star \\
\end{array}\right) =
\left(\begin{array}{cc}
M_{11} & M_{12} \\
M_{21} & M_{22} 
\end{array}\right)
\left(\begin{array}{c}
w^\star \\
\theta_w^\star \\
\end{array}\right) \;  \; ,
\end{equation}
where the elements of the transport matrix can be calculated from the 
optical function $\beta$ and its derivative with respect to $z$ and $\Psi$. 
The transport matrix {\bf M} must be calculated separately in $x$ and $y$ 
and depends on the longitudinal position $z$; the corresponding indices have been 
dropped for clarity. 
While the focusing properties of the beam optics in the vertical plane enable 
a reconstruction of the scattering angle using only $M_{12}$ 
with good precision, the phase advance in the horizontal plane is 
close to $180^{\circ}$ and different reconstruction methods are investigated.

The ALFA detector was designed to use the ``subtraction'' method, exploiting the 
fact that for elastic scattering the particles are back-to-back, that the scattering angle at the A- and C-sides 
are the same in magnitude and opposite in sign, and that the protons originate from the same vertex. 
The beam optics was optimized to maximize the lever arm  $M_{12}$ in the vertical plane in order to access the  
smallest possible scattering angle. The positions measured with ALFA at the A- and C-side 
of ATLAS are roughly of the same size but opposite sign and in the subtraction method the 
scattering angle is calculated according to:
\begin{equation}
\label{eq:subtraction}
\theta^\star_w = \frac{w_\mathrm{A} - w_\mathrm{C}}{M_{12,\mathrm{A}} + M_{12,\mathrm{C}}} \; \; .
\end{equation}
This is the nominal method in both planes and yields the best $t$-resolution. 
An alternative method for the reconstruction of the horizontal scattering angle is to use the 
``local angle'' $\theta_w$ measured by the two detectors on the same side :
\begin{equation}
\label{eq:localangle}
\theta^\star_w = \frac{\theta_{w,\mathrm{A}} - \theta_{w,\mathrm{C}}}{M_{22,\mathrm{A}} + M_{22,\mathrm{C}}} \; \; .
\end{equation}
Another method performs a ``local subtraction'' of measurements at the inner station 
at 237 m and the outer station at 241 m, separately at the A- and C-side, before combining the two sides:
\begin{equation}
\theta^\star_{w,S} = \frac{M_{11,S}^{241} \times w_{237,S} - M_{11,S}^{237} \times w_{241,S}}{M_{11,S}^{241} \times M_{12,S}^{237} - M_{11,S}^{237} \times M_{12,S}^{241}} \; \; , \; S=\mbox{A, C} \; \; .
\end{equation}
Finally, the ``lattice'' method uses both the measured positions and the local angle to reconstruct the scattering angle 
by the inversion of the transport matrix
\begin{equation}\label{eq:lattice_gen}
\left(\begin{array}{c}
w^\star \\
\theta_w^\star 
\end{array}\right) =
{\bf M^{-1}} 
\left(\begin{array}{c}
w \\
\theta_w \\
\end{array}\right) \; , 
\end{equation}
and from the second row of the inverted matrix the scattering angle is determined 
\begin{equation}
\theta^\star_{w} = M_{12}^{-1} \times w + M_{22}^{-1} \times \theta_w \; \; .
\end{equation}
All methods using the local angle suffer from a poor resolution due to a moderate 
angular resolution of about 10 $\mu$rad. Nevertheless, these alternative methods are used to 
cross-check the subtraction method and determine beam optics parameters.  

For all methods $t$ is calculated from the scattering angles as follows:
\begin{eqnarray}
- t & = &\left((\theta^\star_{x})^2 + (\theta^\star_{y})^2  \right)p^2   \; \; ,
\end{eqnarray}
where $\theta^\star_{y}$ is always reconstructed with the subtraction method, because of the parallel-to-point focusing 
in the vertical plane, while all four methods are used for $\theta^\star_{x}$. Results on $\sigmatot$ using 
the four methods are discussed in Section~\ref{sec:sigma_tot}.

\section{Theoretical prediction and Monte Carlo simulation}
\label{sec:mc_th}
Elastic scattering is related to the total cross section through the optical theorem (Eq.~\eqref{eq:OpticalTheorem}) 
and the differential elastic cross section is obtained from the scattering amplitudes of the 
contributing diagrams:
\begin{equation}
\label{eq:elamplitudes}
\frac{\mathrm{d}\sigma}{\mathrm{d}t} = \frac{1}{16\pi}\left|f_{\mathrm{N}}(t) + f_{\mathrm{C}}(t)\mathrm{e}^{\mathrm{i}\alpha\phi(t)}\right|^2 \; \; .
\end{equation}
Here, $f_{\mathrm{N}}$ is the purely strongly interacting amplitude, $f_{\mathrm{C}}$ is the Coulomb amplitude and 
a phase $\phi$ is induced by long-range Coulomb interactions \cite{Bethe,WestAndYennie}.
The individual amplitudes are given by 
\begin{eqnarray}\label{eq:twoamplitudes}
 f_{\mathrm{C}}(t) & = & -8\pi\alpha\hbar c\frac{G^2(t)}{|t|} \;\; , \\
 f_{\mathrm{N}}(t) & = & \left(\rho + \mathrm{i}\right)\frac{\sigmatot}{\hbar c}\mathrm{e}^{-B|t|/2} \;\; , 
\end{eqnarray}
where $G$ is the electric form factor of the proton, $B$ the nuclear slope and 
$\rho=\mathrm{Re}(f_{\mathrm{el}}) / \mathrm{Im}(f_{\mathrm{el}})$. 
The expression for the nuclear amplitude $f_{\mathrm{N}}$ is an approximation valid at small $|t|$ only. 
This analysis uses the calculation of the Coulomb phase from Ref.~\cite{WestAndYennie}
with a conventional dipole parameterization of the proton electric form factor from Ref.~\cite{Cahn}.
The theoretical form of the $t$-dependence of the cross section is obtained from 
the evaluation of the square of the complex amplitudes: 
\begin{eqnarray}\label{eq:tgen}
\frac{\mathrm{d}\sigma}{\mathrm{d}t} & = & \frac{4\pi\alpha^2(\hbar c)^2}{| t |^2} \times G^4(t) \\ \nonumber
 & - &  \; \; \sigmatot \times \frac{\alpha G^2(t)}{|t|}\left[\sin\left(\alpha\phi(t)\right) + \rho \cos\left(\alpha\phi(t)\right) \right] \times \exp{\frac{-B| t |}{2}} \\ \nonumber
 & + &  \; \; \sigmatot^2 \frac{1+\rho^2}{16\pi(\hbar c)^2} \times \exp\left({-B| t |}\right) \; \; ,
\end{eqnarray}
where the first term corresponds to the Coulomb interaction, the second to the Coulomb--nuclear interference 
and the last to the hadronic interaction. This parameterization is used to fit the differential elastic cross 
section to extract $\sigmatot$ and $B$. The inclusion of the Coulomb interaction in the fit of the total cross 
section increases the value of $\sigmatot$ by about $0.6$~mb, compared with a fit with the nuclear term only.

The value of $\rho$  is extracted from global 
fits performed by the COMPETE Collaboration to lower-energy elastic-scattering data 
comprising results from a variety of initial states \cite{compete,PDG_2005}. 
Systematic uncertainties originating 
from the choice of model are important and are addressed e.g.\ in Ref.~\cite{cudell_dispersion}. 
Additionally, 
the inclusion of different data sets in the fit influences the value of $\rho$, 
as described in Refs.~\cite{PDG_2012,menon_silva}. 
In this paper the value from Ref.~\cite{compete} is used with a conservative estimate of the 
systematic uncertainty to account for the model dependence: $\rho=0.140\pm0.008$.  
The theoretical prediction given by Eq.~\eqref{eq:tgen} also depends on 
the Coulomb phase $\phi$ and the form factor $G$. Uncertainties in the Coulomb phase are estimated by 
replacing the simple parameterization from Ref.~\cite{WestAndYennie} with alternative calculations from 
Refs.~\cite{Cahn} and \cite{KFK}, which both predict a different $t$-dependence of the phase. Changing this 
has only a minor fractional impact of order $0.01$~mb on the cross-section prediction. 

The uncertainty on the electric form factor is derived from a comparison of the simple 
dipole parameterization used in this analysis to more sophisticated forms \cite{Simon_FF}, 
which describe high-precision low-energy elastic electron--proton data better \cite{A1}. 
Replacing the dipole by other forms also has a negligible impact of order $0.01$~mb on the 
total cross-section determination. 

Alternative parameterizations of the nuclear amplitude which deviate from the simple exponential $t$-dependence 
are discussed in Section~\ref{sec:model_dependence}.  

\subsection{Monte Carlo simulation}
\label{sec:sim} 
Monte Carlo simulated events are used to calculate acceptance and unfolding corrections.
The generation of elastic-scattering events 
is performed with PYTHIA8 \cite{PYTHIA,PYTHIA6} version 8.165, 
in which the $t$-spectrum is generated according to Eq.~\eqref{eq:tgen}. The divergence of the incoming 
beams and the vertex spread are set in the simulation according to the measurements described in 
Section~\ref{sec:data_taking}.  
After event generation, the elastically scattered protons are transported from the interaction point 
to the RPs, either by means of the transport matrix Eq.~\eqref{eq:transport} or 
by the polymorphic tracking code  
module for the symplectic thick-lens tracking implemented in the MadX~\cite{madx} beam optics calculation package.

A fast parameterization of the detector response is used for the detector simulation with  
the detector resolution tuned to the measured resolution. The resolution is measured by extrapolating 
tracks reconstructed in the inner stations to the outer 
stations using beam optics matrix-element ratios and comparing predicted positions with measured 
positions. It is thus a convolution of the resolutions in the inner and outer stations. 
The fast simulation is tuned to reproduce this convolved resolution. 
A full GEANT4~\cite{GEANT41,GEANT42} simulation is used to set the resolution scale between detectors at the inner and 
outer stations, which cannot be determined from the data. The resolution of the detectors at the outer stations is slightly worse 
than for the detectors at the inner stations as multiple scattering and shower fragments from the latter degrade the 
performance of the former. 

Systematic uncertainties from the resolution difference between the detectors at the inner and outer stations  
are assessed by fixing the resolution of the detectors  at the inner stations either to the value from GEANT4 or to 
the measurement 
from the test beam~\cite{testbeam1,testbeam2} and matching the resolution at the outer stations to reproduce the 
measured convolved resolution. The resolution depends slightly on the vertical track position on the detector 
surface; a further systematic uncertainty is assessed by replacing the constant resolution by a linear parameterization.  
The total systematic uncertainty on the resolution of about $\pm 10\%$ is dominated by the difference between the value from the test-beam 
measurement and from GEANT4 simulation.

\section{Data taking}
\label{sec:data_taking}
The data were recorded in a dedicated low-luminosity run using beam optics with a $\betastar$ of 90 m; 
details of the beam optics settings can be found in Refs.~\cite{Note90m,HBLumiday11}. 
The duration of this run was four hours. 
For elastic-scattering events, the main pair of colliding bunches was used, which contained around 7$\times$10$^{10}$ protons per bunch.
Several pairs of pilot bunches with lower intensity and unpaired bunches were used for the studies of systematic
uncertainties. 
The rates of the head-on collisions were maximized using measurements from online luminosity monitors.
The run used for data analysis was preceded by a run with identical beam optics to align the RPs.   

\subsection{Beam-based alignment}
\label{sec:Fill_Scraping}
Very precise positioning of the RPs is mandatory to achieve the 
desired precision on the position measurement of 20--30~$\mu$m in both the horizontal and vertical 
dimensions. 
The first step is a beam-based alignment procedure to determine the position of the RPs with 
respect to the proton beams. One at a time, 
the eight pots are moved into the beam, ultimately by steps of 10~$\mu$m only, until the LHC beam-loss monitors give a 
signal well above threshold. 
The beam-based alignment procedure was performed in a dedicated fill with identical beam settings just before the data-taking run. 
The vertical positions of the beam envelopes were determined at each station by scraping the beam with the upper and lower RPs in turn.
\par
From the positions of the upper and the lower RP windows with respect to the beam edges, the centre of the beam as well as the distance between
the upper and lower pots were computed. 
For the two stations on side C the centres were off zero by typically 0.5--0.6 mm; for side A both were off by about $-0.2$ mm.
On each side, the distances between the upper and the lower pots
were measured to be $8.7$ mm for the station nearer to the interaction point and $7.8$ mm for the far one. This difference corresponds to the change
in the nominal vertical beam widths ($\sigma_{y}$) between the two stations. 
\par
The data were collected with the pots at 6.5$\times\sigma_{y}$ from the beam centre, the closest possible distance 
with reasonable background rates. 
With a value of the nominal vertical beam spread,  $\sigma_{y}$, of 897 $\mu$m (856 $\mu$m) for the inner (outer) stations, 
the 6.5$\times\sigma_{y}$ positions correspond to a typical distance of 5.83~mm (5.56~mm) from the beam line in the LHC reference frame.

\subsection{Beam characteristics: stability and emittance}
\label{sec:beam_conditions}
Beam position monitors, regularly distributed along the beam line 
between the interaction point and the RPs, were used to survey the horizontal and vertical positions of the beams. 
The variations in position throughout the duration of the data taking were of the order of 10 $\mu$m 
in both directions, which is equal to the precision of the measurement itself.

The vertical and horizontal beam emittances $\epsilon_{y}$ and $\epsilon_{x}$, expressed in $\mu$m, are used in 
the simulation to track a scattered proton going from the interaction point
to the RPs and therefore to determine the acceptance. These were measured at regular intervals during the fill using a 
wire-scan method~\cite{Wire_Scan} and monitored bunch-by-bunch throughout the run using 
two beam synchrotron radiation monitor systems; the latter were calibrated to the wire-scan measurements at the start of the run.
The emittances varied smoothly from 2.2~$\mu$m to 3.0~$\mu$m (3.2~$\mu$m to 4.2~$\mu$m) for 
beam 1 (beam 2) in the horizontal plane and  
from 1.9 $\mu$m to 2.2 $\mu$m (2.0~$\mu$m to 2.2~$\mu$m) for beam 1 (beam 2) in the vertical plane. The systematic 
uncertainty on the emittance is about $10\%$.   
A luminosity-weighted average emittance is used in the simulation, resulting in an angular beam divergence of about 3 $\mu$rad. 

\subsection{Trigger conditions}
\label{sec:trigger}
To trigger on elastic-scattering events, two main triggers were used. 
The triggers 
required a coincidence of the main detector trigger scintillators between either
of the two upper (lower) detectors on side A and either of the two lower (upper) detectors on side C. 
The elastic-scattering rate was typically 50~Hz in each arm.
The trigger efficiency for elastic-scattering events was determined from a data stream in which all events with a hit in 
any one of the ALFA trigger counters were recorded. In the geometrical acceptance of the detectors, the efficiency of the trigger used to record 
elastic-scattering events is $99.96 \pm 0.01\%$.

\section{Track reconstruction and alignment}
\label{sec:tracking}
\subsection{Track reconstruction}
\label{sec:track_reco}
The reconstruction of elastic-scattering events is based on local tracks 
of the proton trajectory in the RP stations. 
A well-reconstructed elastic-scattering event consists of local tracks in all four RP stations.    

The local tracks in the MDs are reconstructed from the hit pattern of 
protons traversing the scintillating fibre layers.
In each MD, 20 layers of scintillating fibres are arranged perpendicular to the 
beam direction. 
The hit pattern of elastically scattered protons is characterized by a straight trajectory, almost 
parallel to the beam direction.  
In elastic events the average multiplicity per 
detector is about 23 hits, where typically 18--19 are attributed to the proton trajectory 
while the remaining 4--5 hits are due to beam-related background, cross-talk and electronic noise.

\begin{figure}[h!]
  \centering
  \includegraphics[width=140mm]{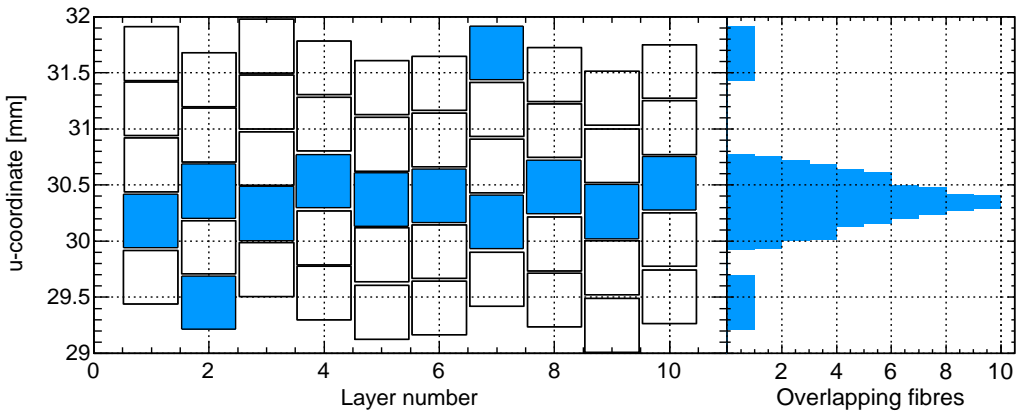}
  \caption{Hit pattern of a proton trajectory in the ten fibre layers comprising the 
   $u$ coordinate. The superposition of fibre hits attributed to a track is shown in 
   the histogram. The position of maximum overlap is used to determine the track position.}
\label{fig:Event_Disp}
\end{figure}

The reconstruction assumes that the protons pass through the fibre detector perpendicularly.
A small angle below 1 mrad with respect to the beam direction has no sizeable impact. 
The first step of the reconstruction is to determine  
the $u$ and $v$ coordinates from the two sets of ten layers 
which have the same orientation.
The best estimate of the track position is given by the overlap region of the fiducial 
areas of all hit fibres. As illustrated in Fig.~\ref{fig:Event_Disp}, the staggering 
of the fibres narrows the overlap region and thereby improves the resolution.    
The centre of the overlap region gives the $u$ or $v$ coordinate, while the width 
determines the resolution.
Pairs of $u$ and $v$ coordinates are transformed to spatial positions in the 
beam coordinate system.

To exclude events with hadronic showers and layers with a high noise level, fibre layers 
with more than ten hits are not used in track reconstruction.
At least three layers out of the possible ten are required to have a hit multiplicity between one 
and three.
Finally, the $u$ and $v$ coordinates accepted for spatial positions must be formed from
at least three overlapping hit fibres. 

If more than one particle passes through the detector in a single event, the orthogonal 
fibre geometry does not allow a unique formation of tracks.
In elastic-scattering events,
multiple tracks can originate from associated halo or shower particles  
as well as the overlap of two events. 
The first type of multiple tracks happens mostly at one side of the spectrometer arms 
and can be removed by track-matching with the opposite side. 
In the case of elastic pile-up, where multiple tracks are reconstructed in both arms, 
only the candidate 
with the best track-matching is accepted. The fraction of genuine pile-up is about  
0.1\% and is corrected by a global factor as described in Section~\ref{sec:event_selection}. 

The reconstruction of tracks in the ODs 
is based on the same method as described here for the MDs, but with reduced precision 
since only three fibre layers are available.

\subsection{Alignment}
\label{sec:alignment}
The precise positions of the 
tracking detectors with respect to the circulating beams are crucial inputs 
for the reconstruction of the proton kinematics.
For physics analysis, the detector positions  
are directly determined from the elastic-scattering data.

The alignment procedure is based on the distribution of track positions in
the RP stations in the full elastic-scattering event sample. This distribution forms a
narrow ellipse with its major axis in the vertical ($y$) direction, with an aperture gap
between the upper and lower detectors.
The measured distances between upper and lower MDs and the 
rotation symmetry of scattering angles are used as additional constraints.  

Three parameters are necessary to align each MD: the horizontal and 
vertical positions and the rotation angle around the beam axis. 
A possible detector rotation around the horizontal or vertical axes can be of the 
order of a few mrad, as deduced from survey measurements and the alignment 
corrections for rotations around the beam axis.
Such tiny deviations from the nominal angles result in 
small offsets which are effectively absorbed in the three alignment parameters.

The horizontal detector positions and the rotation angles are determined from a fit of a straight 
line to a profile histogram of the narrow track patterns in the upper and lower MDs.     
The uncertainties are 
1--2~$\mu$m for the horizontal coordinate and 0.5 mrad for the angles.

For the vertical detector positioning, the essential input is the distance between the upper 
and lower MDs. Halo particles which pass the upper and lower ODs at the same time are used 
for this purpose. Combining the two $y$-coordinates allows this distance 
to be determined. 
Many halo events are used to improve the precision of the distance value by averaging over 
large samples. The associated systematic uncertainty in the distance value is derived from variations 
of the requirements on hit and track multiplicities in the ODs.    
The measured distances between the upper and lower MDs for the nominal position at 
6.5$\times\sigma_{y}$ and related errors are
shown in Table~\ref{tab:vert_alignment}.
\begin{table}
  \begin{center}
    \begin{tabular}{llcccccc}
      \hline \hline
      Detector & Station & Distance [mm] & $y_{\rm{meas}}$  [mm]& $y_{\rm{ref}}$ [mm] \\
      \hline
       A1 & \multirow{2}{*}{B7L1} & $11.962\pm0.081$ & $\phantom{+}5.981\pm0.093$ & $\phantom{+}5.934\pm0.076$ \\
       A2 &  & $11.962\pm0.081$ &$-5.981\pm0.093$ &$-5.942\pm0.076$ \\
       A3 & \multirow{2}{*}{A7L1} & $12.428\pm0.022$ & $\phantom{+}6.255\pm0.079$ & $\phantom{+}6.255\pm0.078$ \\
       A4 &  & $12.428\pm0.022$ &$-6.173\pm0.079$ &$-6.173\pm0.080$ \\
       A5 &  \multirow{2}{*}{A7R1} & $12.383\pm0.018$ & $\phantom{+}6.080\pm0.078$ & $\phantom{+}6.036\pm0.088$ \\
       A6 &  &  $12.383\pm0.018$ &$-6.304\pm0.078$ &$-6.273\pm0.087$ \\
       A7 & \multirow{2}{*}{B7R1} & $11.810\pm0.031$ & $\phantom{+}5.765\pm0.077$ & $\phantom{+}5.820\pm0.083$ \\
       A8 &  & $11.810\pm0.031$ &$-6.045\pm0.077$ &$-6.120\pm0.084$ \\
      \hline \hline
    \end{tabular}
    \caption{The distance between the upper and lower MDs and related vertical detector
             positions. The coordinates $y_{\rm{meas}}$ refer to the individual alignment per
             station while the coordinates $y_{\rm{ref}}$ are derived by extrapolation
             using beam optics from the detector positions in the reference station A7L1.
             The quoted errors include statistical and systematic contributions.}
  \label{tab:vert_alignment}
  \end{center}
\end{table}
The main contributions to the uncertainty on the distance are due to uncertainties in the 
fibre positions and the relative alignment of the ODs with respect to the MDs, both  
typically 10 $\mu$m.
The most precise distance values, with uncertainties of about 20 $\mu$m, are
achieved for the two inner stations A7L1 and A7R1. The values for the outer stations 
are degraded by shower particles from interactions in the inner stations. The large 
error on the distance in station B7L1 is due to a detector which was not calibrated in a test beam. 

The absolute vertical detector positions with respect to the beam are determined using the criterion of equal track 
densities in the upper and lower MDs at identical distances from the beam.
The reconstruction efficiency in each spectrometer arm, as 
documented in Section~\ref{sec:reco_efficiency}, is taken into account. 
A sliding window technique is applied to find the positions 
with equal track densities. 
The resulting values $y_{\text{meas}}$ are summarized in 
Table~\ref{tab:vert_alignment}. 
The typical position error is about 80~$\mu$m, dominated by the uncertainties 
on the reconstruction efficiency. 

For the physics analysis, all vertical detector positions are derived from 
the detector positions in a single reference station by means of the beam optics.
The measured lever arm ratios $M_{12}^{\text{X}}/M_{12}^{\text{A7L1}}$, with X labelling 
any other station, are used for the extrapolation from the reference station to the
other stations. 
As a reference the detector positions $y_{\text{meas}}$ of the 
inner station A7L1 with a small distance error were chosen.
The resulting detector positions $y_{\text{ref}}$ are also given in 
Table~\ref{tab:vert_alignment}.
Selecting the other inner station A7R1 as a reference indicates 
that the systematic uncertainty due to the choice of station is about 20~$\mu$m, 
which is well covered by the total position error.

\section{Data analysis}
\label{sec:data_analysis}

\subsection{Event selection}
\label{sec:event_selection}

All data used in this analysis were recorded in a single run and only events resulting from collisions of the bunch pair with 
about 7$\times$10$^{10}$ protons per bunch  
were selected; events from pilot bunches were discarded.
Furthermore, only periods of the data taking where the dead-time 
fraction was below $5\%$ are used. This requirement eliminates a few minutes of the run. The average dead-time fraction 
was $0.3\%$ for the selected data.

Events are required to pass the trigger conditions for elastic-scattering events, 
and have a reconstructed track in all four detectors of the
arm which fired the trigger. 
Events with additional tracks in  
detectors of the other arm arise from the overlap of halo protons with 
elastic-scattering protons and are retained. In the case of an overlap of halo and elastic-scattering 
protons in the same detectors, 
which happens typically only on one side,
a matching procedure between the detectors on each side is applied to identify the elastic-scattering track.  

\begin{figure}[h!]
  \centering
  \includegraphics[width=100mm]{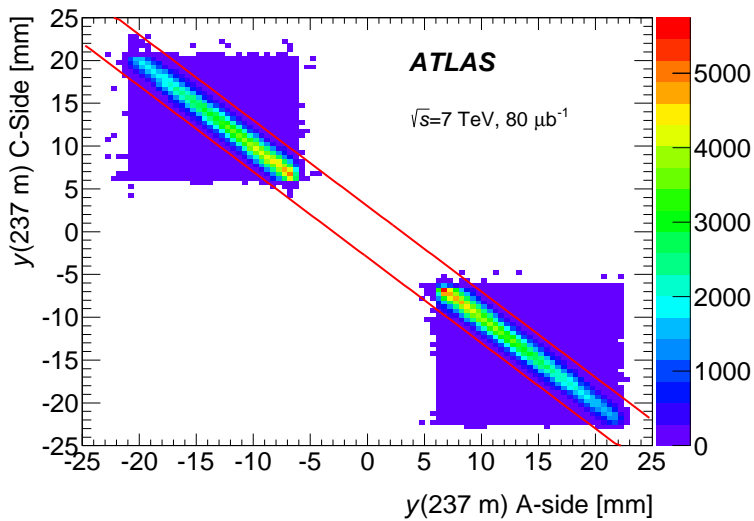}
  \caption{The correlation of the $y$ coordinate measured at the A- and C-side for the inner stations. 
Elastic-scattering candidates after data quality, trigger and bunch selection but before 
acceptance and background rejection cuts are shown. Identified elastic events are required to lie between the red lines.} 
  \label{fig:evsel_y_left_right}
\end{figure}

\begin{figure}[h!]
  \centering
  \includegraphics[width=100mm]{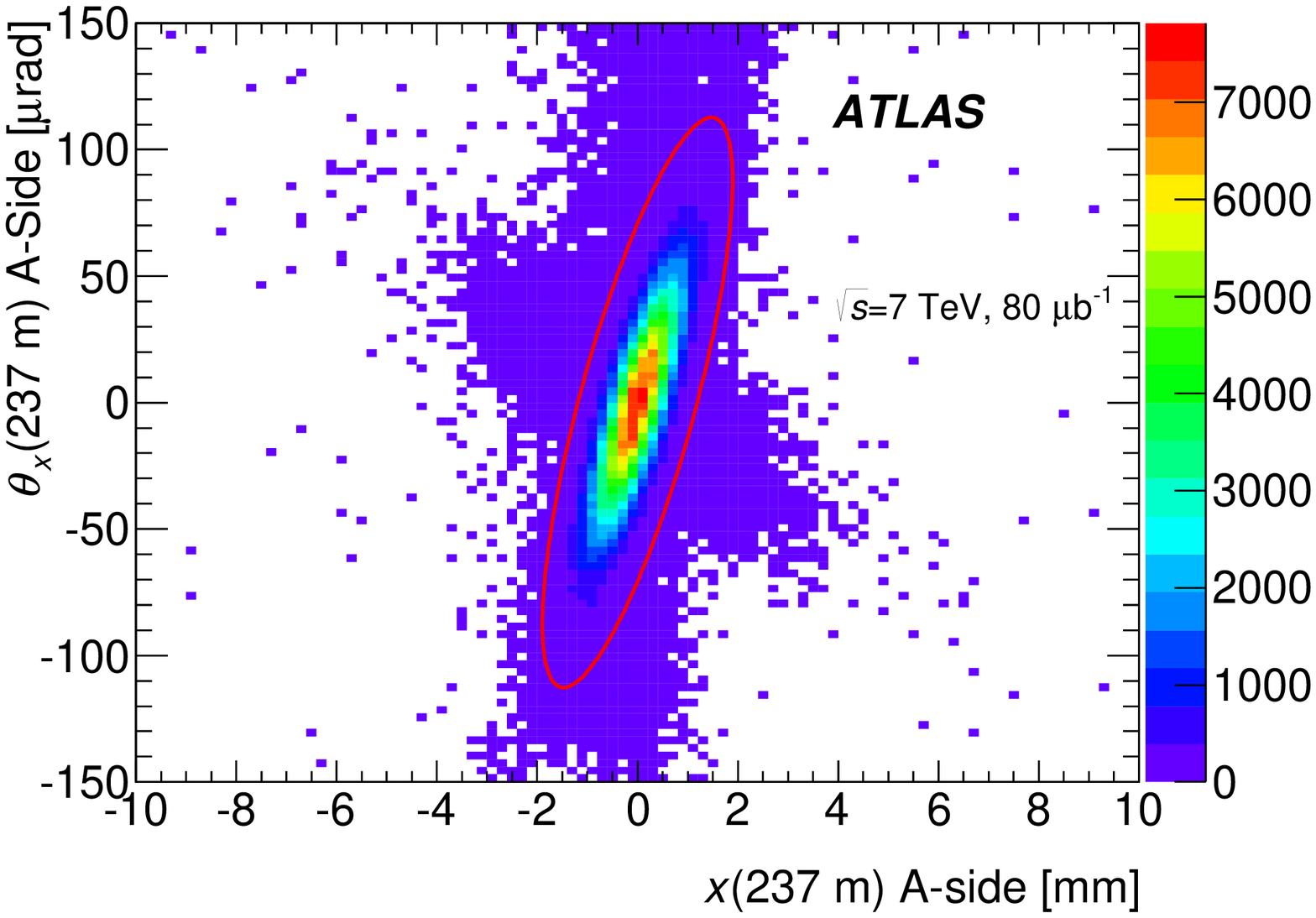}
  \caption{The correlation between the horizontal coordinate, $x$, and the local horizontal angle, $\theta_x$, on the A-side.  
Elastic-scattering candidates after data quality, trigger and bunch selection but before 
acceptance and background rejection cuts are shown. Identified elastic events are required to lie inside the red ellipse.} 
  \label{fig:evsel_x_thetax}
\end{figure}

Further geometrical cuts on the left-right acollinearity are applied, exploiting the back-to-back topology of 
elastic-scattering events. The position difference between the left and the right sides is required to be within 3.5$\sigma$ 
of its resolution determined from simulation, as shown in Fig.~\ref{fig:evsel_y_left_right} for the 
vertical coordinate. An efficient cut against non-elastic background is obtained from 
the correlation of the local angle between two stations and the position in the horizontal plane, as shown in 
Fig.~\ref{fig:evsel_x_thetax}, where elastic-scattering events appear inside a narrow ellipse with positive slope, whereas 
beam-halo background is concentrated in a broad ellipse with negative slope or in an uncorrelated band.  

Finally, fiducial cuts to ensure a good containment inside the detection area are applied to the 
vertical coordinate. It is required to be at least 60 $\mu$m from the edge of the detector nearer the beam, where the full detection efficiency 
is reached. At large vertical distance,  the vertical coordinate must be at least 1 mm away from the shadow 
of the beam screen, a protection 
element of the quadrupoles, in order to minimize the impact from showers generated in the beam screen.

The numbers of events in the two arms after each selection criterion  
are given in Table~\ref{tab:cutflow}. 
 \begin{table}
  \begin{center}
    \begin{tabular}{lcccc}
      \hline \hline
Selection criterion     &  \multicolumn{4}{c}{Numbers of events} \\ 
      \hline
Raw number of events    & \multicolumn{4}{c}{6620953}  \\
Bunch group selection   & \multicolumn{4}{c}{1898901}  \\
Data quality selection  & \multicolumn{4}{c}{1822128}  \\
Trigger selection       & \multicolumn{4}{c}{1106855}  \\
                        & Arm 1 & fraction & Arm 2 & fraction \\
Reconstructed tracks     & 459229 & & 428213 & \\
Cut on $x$ left vs right & 445262 & $97.0\%$ & 418142 & $97.6\%$\\
Cut on $y$ left vs right & 439887 & $95.8\%$ & 414421 & $96.8\%$\\
Cut on $x$ vs $\theta_x$ & 434073 & $94.5\%$ & 410558 & $95.9\%$\\
Beam-screen cut          & 419890 & $91.4\%$ & 393320 & $91.9\%$\\
Edge cut                 & 415965 & $90.6\%$ & 389463 & $91.0\%$\\ \hline
Total selected & \multicolumn{4}{c}{805428} \\
Elastic pile-up & \multicolumn{4}{c}{1060} \\
     \hline \hline
    \end{tabular}
  \caption{Numbers of events after each stage of the selection. The fractions of events surviving the 
event-selection criterion with respect to the total number of reconstructed events are shown for each criterion.  
The last row gives the number of observed pile-up events.}
  \label{tab:cutflow}
  \end{center}
\end{table}
At the end of the selection procedure 805,428 events survive all cuts. 
A small asymmetry is observed between the two arms, which arises from the detectors 
being at different vertical distances, asymmetric beam-screen positions and background 
distributions. A small number of elastic pile-up events corresponding to a $0.1\%$ fraction is observed, where two elastic events 
from the same bunch crossing are reconstructed in two different arms. 
Elastic pile-up events can also occur with two protons in the same detectors; in this 
case only one event is counted as described in Section~\ref{sec:track_reco}. A correction is applied for the pile-up events appearing 
in the same arms, by scaling the pile-up events in different arms by a factor of two.

\subsection{Background determination}
\label{sec:background}
A small fraction of the background is expected to be inside the area defined 
by the elliptical contour used to reject 
background shown in Fig.~\ref{fig:evsel_x_thetax}. 
The background events peak  
at small values of $x$ and $y$ and thus constitute an irreducible background at small $|t|$. 
The background predominantly originates from accidental coincidences of beam-halo particles, but  
single diffractive protons in coincidence with a halo proton at the opposite side may also contribute. 

\begin{figure}[h!]
  \centering
  \includegraphics[width=100mm]{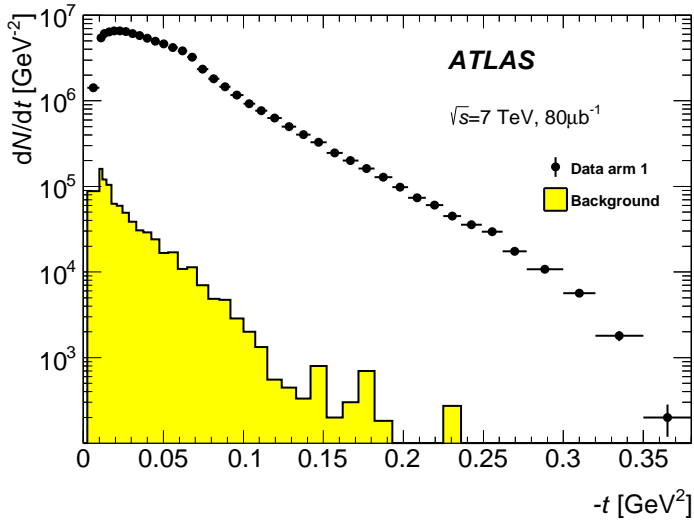}
  \caption{The counting rate $\mathrm{d}N/\mathrm{d}t$, before corrections, as a function of $t$ in arm 1 compared to the background spectrum 
determined using anti-golden events. The form of the distribution is modified by acceptance 
effects (see Fig.~\ref{fig:acceptance}).} 
\label{fig:background_antigolden}
\end{figure}

While elastic-scattering events are selected in the ``golden'' topology with two tracks in opposite 
vertical detector positions on the left and right side, events in the ``anti-golden'' topology 
with two tracks in both upper or both lower detectors at the left and right side are pure 
background from accidental coincidences. After applying the event selection cuts, these events yield an estimate 
of background in the elastic sample with the golden topology. 
Furthermore, the anti-golden events can be used to calculate the form of the $t$-spectrum for background 
events by flipping the sign of the vertical coordinate on either side. As shown in Fig.~\ref{fig:background_antigolden}, 
the background $t$-spectrum peaks strongly at small $t$ and falls off steeply, distinguishably different from 
the distribution obtained for elastic events. 

Alternatively, the amount of background per arm is determined from the distribution of the horizontal 
vertex position at the interaction point, reconstructed using the beam optics transport Eq.~\eqref{eq:transport}. For elastic scattering,  
the vertex position peaks at small values of $x$, whereas for background the shape is much broader since 
halo background does not originate from the interaction point. Hence the fraction of background events can be determined 
from a fit to the measured distribution using templates. 

The anti-golden method is used to estimate the background. Systematic uncertainties 
in the normalization are taken from the difference between the anti-golden method and 
the vertex method, 
while the background shape uncertainty is obtained from variations of the flipping procedure used to 
transform the anti-golden events into elastic-like events. 
 \begin{table}
  \begin{center}
    \begin{tabular}{lcc}
      \hline \hline
     &  Arm $++$ & Arm $--$ \\ 
      \hline
Numbers of background events  & 3329 & 1497   \\
Statistical error & $\pm 58$ & $\pm 39$ \\
Systematic error  & $\pm 1100$ & $\pm 1200$ \\
     \hline \hline
    \end{tabular}
  \caption{Number of background events in each arm estimated with the anti-golden method 
  and systematic uncertainties from the difference from the vertex method. The arm $++$ comprises all four upper detectors, 
  the arm $--$ all four lower detectors.}
  \label{tab:antigolden}
  \end{center}
\end{table}
The expected numbers of background events are given in Table~\ref{tab:antigolden} together with their uncertainties.  
The total uncertainty on the background is dominated by the systematic uncertainty of 50--80\%.  
Given the overall small background 
contamination of about $0.5\%$, the large systematic uncertainty has only a small impact on the total cross-section 
determination.

\subsection{Event reconstruction efficiency}
\label{sec:reco_efficiency}
Elastic-scattering events inside the acceptance region are expected to have a proton track in each of the four detectors of the corresponding spectrometer arm.
However, in the case of interactions of the protons or halo particles with the stations or detectors, which result in too large fibre hit multiplicities, the track reconstruction described in Section~\ref{sec:track_reco} may fail. 
The rate of elastic-scattering events has to be corrected for losses due to such partly reconstructed events.
This correction is defined as the event reconstruction efficiency. 

A method based on a tag-and-probe approach is used to estimate the efficiency. 
Events are grouped into several reconstruction cases, for which different selection criteria and corrections are applied, to determine if an event is from elastically scattered protons, but was not fully reconstructed because of inefficiencies.

The reconstruction efficiency of elastic-scattering events is defined as
$\varepsilon_{\text{rec}} = N_{\text{reco}} / (N_{\text{reco}}+N_{\text{fail}})$,
where $N_{\text{reco}}$ is the number of fully reconstructed elastic-scattering events, which have at least one 
reconstructed track in each of the four detectors of an spectrometer arm, and $N_{\text{fail}}$ is the number of not fully 
reconstructed elastic-scattering events which have reconstructed tracks in fewer than four detectors. 
It does not include the efficiency of other selection cuts, which is discussed in Section \ref{sec:unfolding_acceptance}, and is separate 
from any acceptance effects.
Events of both classes need to have an elastic-scattering trigger signal and need to be inside the acceptance region, i.e.\ they have to fulfill the event selection criteria for elastic-scattering events. 
The efficiency is determined separately for the two spectrometer arms.
Based on the number of detectors with at least one reconstructed track the events are grouped into six reconstruction cases 4/4, 3/4, 2/4, 
(1+1)/4, 1/4 and 0/4. 
Here the digit in front of the slash indicates the number of detectors with at least one reconstructed track. 
In the 2/4 case both detectors with tracks are on one side of the interaction point and in the (1+1)/4 case they are on different sides. 
With this definition one can write
\begin{equation}\label{eq:eff2}
\varepsilon_{\text{rec}} = \frac{N_{\text{reco}}}{N_{\text{reco}}+N_{\text{fail}}} = \frac{N_{4/4}}{N_{4/4}+N_{3/4}+N_{2/4}+N_{(1+1)/4}+N_{1/4}+N_{0/4}},
\end{equation}
where $N_{k/4}$ is the number of events with $k$ detectors with at least one reconstructed track in a spectrometer arm.
The event counts $N_{k/4}$ need to be corrected for background, which is described in the following.

Elastic-scattering events are selected for the various cases based on the event selection criteria described in 
Section \ref{sec:event_selection}. 
The proton with reconstructed tracks on one side of the interaction point is used as a tag and the one on the other side as a probe. 
Both tag and probe have to pass the event selection to be counted as an elastic-scattering event and to be classified as one of the reconstruction cases.
Depending on the case, only a sub-set of the event selection criteria can be used, because just a limited number of detectors with reconstructed tracks is available.
For example it is not possible to check for left-right acollinearity of 2/4 events, because tracks are only reconstructed on one side of the interaction point.
To disentangle the efficiency from acceptance, the total fibre hit multiplicity (sum of all fibre hits in the 20 layers of a detector; maximum is 1280 hits) in all detectors without any reconstructed track has to have a minimum value of six. 
In this way, events where tracks could not be reconstructed due to too few fibre hits are excluded from the efficiency calculation and only handled by the acceptance, which is discussed in Section \ref{sec:unfolding_acceptance}.
\begin{figure}[ht]
  \centering
  \includegraphics[width=100mm]{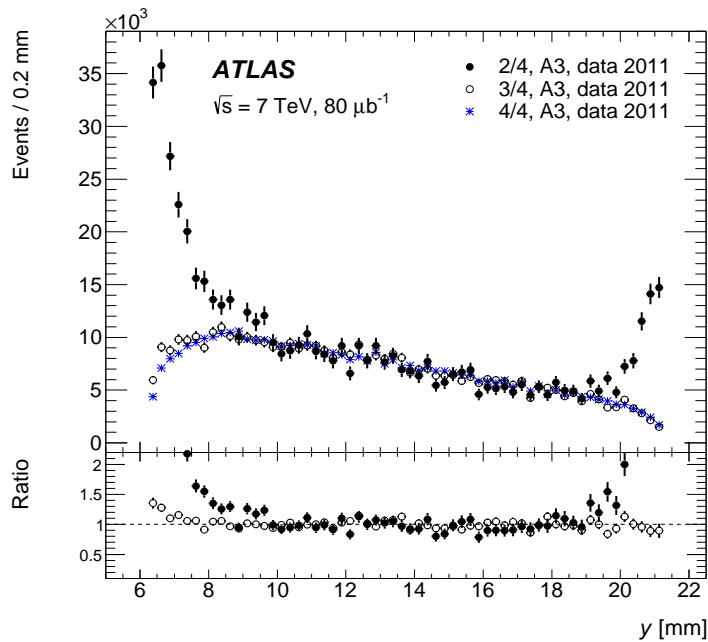}
  \caption{Distribution of the vertical track position at the detector surface for detector A3 of the 2/4 case ($\bullet$) where no track was reconstructed in A6 and A8, the 3/4 case ($\circ$) 
  where no track was reconstructed in A8 and of the 4/4 case ($\textcolor{blue}{\hexstar}$). 
  The distributions are normalized to the number of events in the 4/4 case. 
  The bottom panel displays ratios between $k$/4 and 4/4 with statistical uncertainties.}
  \label{fig:efficiency1}
\end{figure}
\begin{figure}[ht]
  \centering
  \includegraphics[width=100mm]{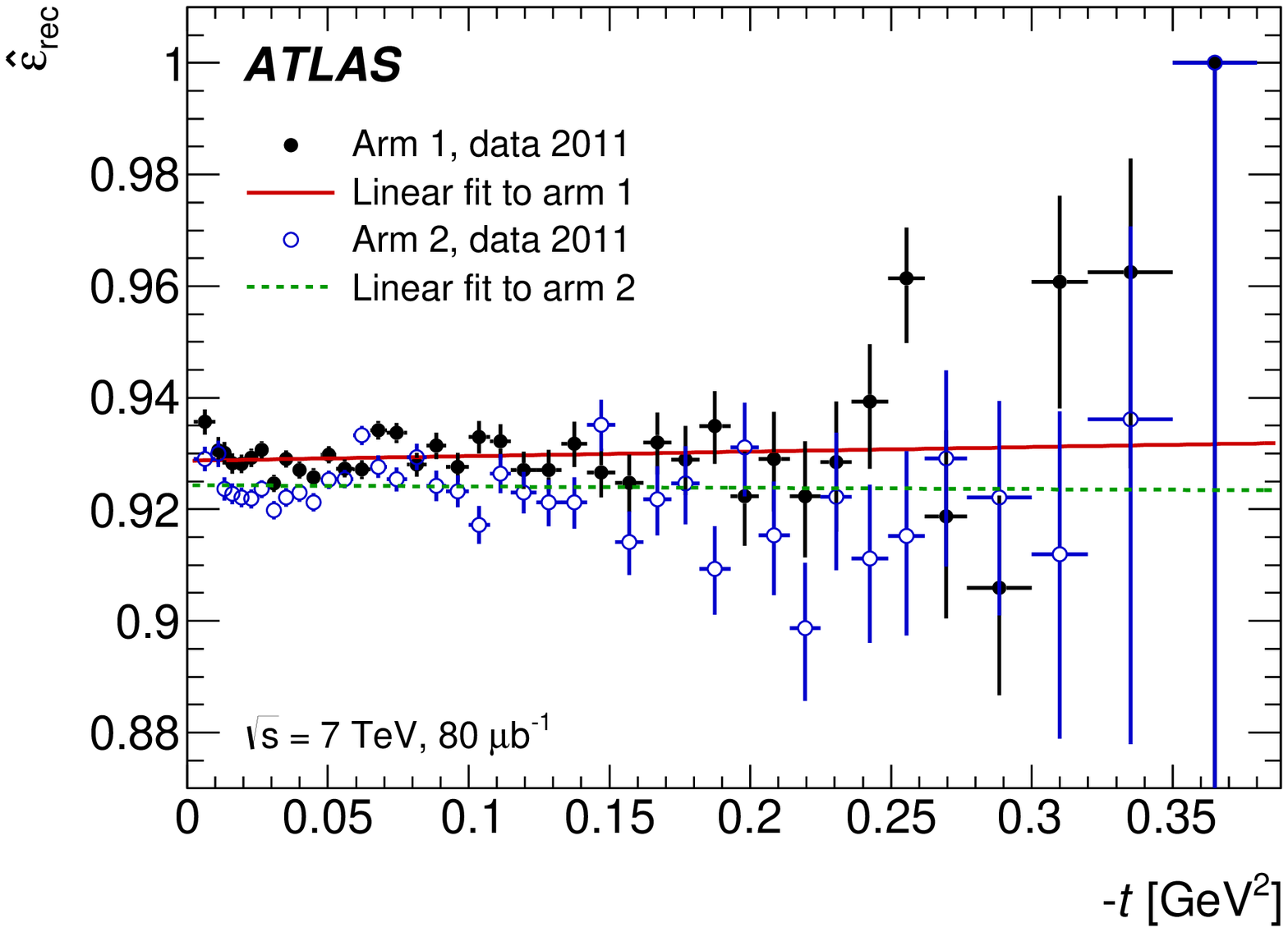}
  \caption{Partial event reconstruction efficiency $\hat{\varepsilon}_\text{rec}$ as a function of $-t$ for elastic arms 1 ($\bullet$) and 2 ($\textcolor{blue}{\circ}$). 
  The solid line is a linear fit to arm 1 and the dashed one to arm 2.}
  \label{fig:efficiency2}
\end{figure}

With events from the 3/4 case it is possible to apply most of the event selection criteria and reconstruct $t$ well with the subtraction method.
The position distribution of reconstructed tracks in 3/4 events agrees very well with that from 4/4 events, as shown in Fig.~\ref{fig:efficiency1} for the vertical coordinate. 
Because of this good agreement and the ability to reconstruct $t$, a partial event reconstruction efficiency $\hat{\varepsilon}_\text{rec}(t) = N_{4/4}(t) / [N_{4/4}(t) + N_{3/4}(t)]$ 
as a function of $t$ can be constructed, as shown in Fig.~\ref{fig:efficiency2}. 
Linear fits, applied to the partial efficiencies of each spectrometer arm, yield small residual slopes consistent with zero within uncertainties. 
This confirms that the efficiency $\varepsilon_\text{rec}$ in Eq.~\eqref{eq:eff2} is independent of $t$, as is expected from the uniform material distribution in the detector volume.

Two complications arise when counting 2/4 events. 
First, the vertical position distributions of the two detectors with reconstructed tracks do not agree with that from 4/4 events. 
As shown in Fig.~\ref{fig:efficiency1} for A3, peaks appear at both edges of the distribution. 
These peaks are caused by events where protons hit the beam screen or thin RP window on one side of the interaction point and tracks are therefore only reconstructed on the other side.
Since these events would be excluded by acceptance cuts, they are removed from the efficiency calculation. 
This is achieved by extrapolating the vertical position distribution from the central region without the peaks to the full region using 
the shape of the distribution from 4/4 events.

The second complication arises from single-diffraction background, which has a similar event topology to the 2/4 events.
This background is reduced by elastic-scattering trigger conditions, but an irreducible component remains. 
Therefore, the fraction of elastic-scattering events is determined with a fit, which uses two templates to fit the horizontal position distributions. 
Templates for elastic-scattering events and a combination of single diffraction and other backgrounds are both determined from data.
For the elastic-scattering template, events are selected as described in Section~\ref{sec:event_selection}. 
Events for the background template are selected based on various trigger signals that enhance the background and reduce the elastic-scattering contribution.
The fit yields an elastic-scattering fraction in the range of $r_\text{el} = $ 0.88 to 0.96, depending on the detector.

Because about 95\% of $N_{\text{fail}}$ consists of 3/4 and 2/4 events, the other cases play only a minor role.
For cases (1+1)/4 and 1/4 an additional event selection criterion on the horizontal position distribution is applied to enhance the contribution from elastic scattering and suppress background. 
Edge peaks are present in the $y$-position distributions of 1/4 events, and the extrapolation procedure, described above, is also applied. 
In the 0/4 case no track is reconstructed in any detector and the event selection criteria cannot be applied. 
Therefore, the number of 0/4 events is estimated from the probability to get a 2/4 event, which is determined from the ratio of the number of 2/4 to the number of 4/4 events.
The contribution to $N_{\text{fail}}$ of this estimated number of 0/4 events is only about 1\%.

The systematic uncertainties on the reconstruction efficiency are determined by varying the event selection criteria. 
Additional uncertainties arise from the choice of central extrapolation region in $y$ for 2/4 and 1/4 events and from the fraction fit. 
The uncertainties on the fraction fit are also determined by selection criteria variation and an additional uncertainty is attributed to the choice of background template.

The event reconstruction efficiencies in arm 1 and arm 2 are determined to be $\varepsilon_{\text{rec,1}} = 0.8974 \pm 0.0004 \stat \pm 0.0061 \syst$ and $\varepsilon_{\text{rec,2}} = 0.8800 \pm 0.0005 \stat \pm 0.0092 \syst$ respectively. 
The efficiency in arm 1 is slightly larger than in arm 2, due to differences in the detector configurations. 
In arm 1 the trigger plates are positioned after the scintillating tracking fibres and in arm 2 they are positioned in front of them. 
This orientation leads to a higher shower probability and a lower efficiency in arm 2.

\section{Acceptance and unfolding}
\label{sec:unfolding_acceptance}

\begin{figure}[h!]
  \centering
  \includegraphics[width=100mm]{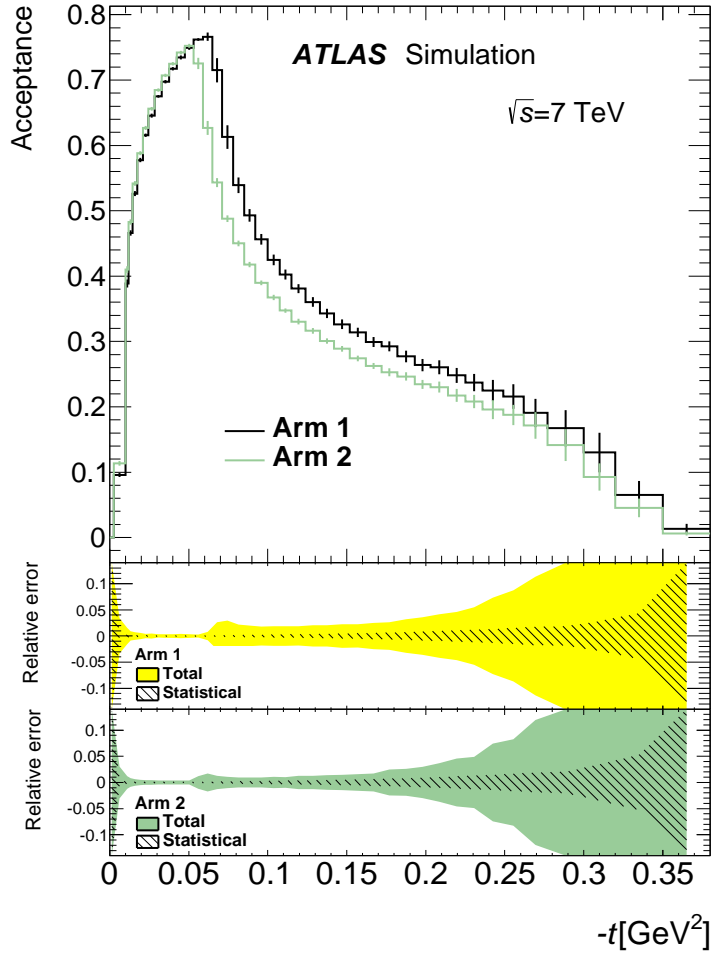}
  \caption{The acceptance as a function of the true value of $t$ for each arm with total uncertainties shown 
as error bars. The lower panels show relative total and statistical uncertainties.} 
  \label{fig:acceptance}
\end{figure}

The acceptance is defined as the ratio of events passing all geometrical and fiducial acceptance cuts 
defined in Section~\ref{sec:event_selection} to all generated events and is calculated as a function of $t$. 
The calculation is carried out with PYTHIA8 as elastic-scattering event generator and MadX for beam transport based 
on the effective optics (see Section~\ref{sec:optics_fit}). 
The acceptance is shown in Fig.~\ref{fig:acceptance} for each arm. 

The shape of the acceptance curve can be understood from the contributions of the vertical and horizontal 
scattering angles to $t$, $-t=((\theta^\star_x)^2 + (\theta^\star_y)^2)p^2$. The smallest accessible value of $t$ is obtained at the detector edge 
and set by the vertical distance of the detector from the beam. Close to the edge, the acceptance is small 
because a fraction of the events is lost due to beam divergence, i.e. events being inside the acceptance on one 
side but outside at the other side. At small $|t|$ up to $-t \sim 0.07 \GeV^2$ vertical and horizontal 
scattering angles contribute about equally to a given value of $t$. Larger $t$-values imply larger 
vertical scattering angles and larger values of $y$, and with increasing $y$ the fraction of events 
lost in the gap between the main detectors decreases. 
The maximum acceptance is reached 
for events occurring at the largest possible values of $y$ within the beam-screen cut. 
Beyond that point the acceptance decreases steadily because the events are
required to have larger values of $x$ since these $t$-values are dominated by the horizontal scattering angle component. 
This also explains 
the difference between the two arms, which is dominated by the difference between the respective beam-screen 
cuts. 

\begin{figure}[h!]
  \centering
  \includegraphics[width=100mm]{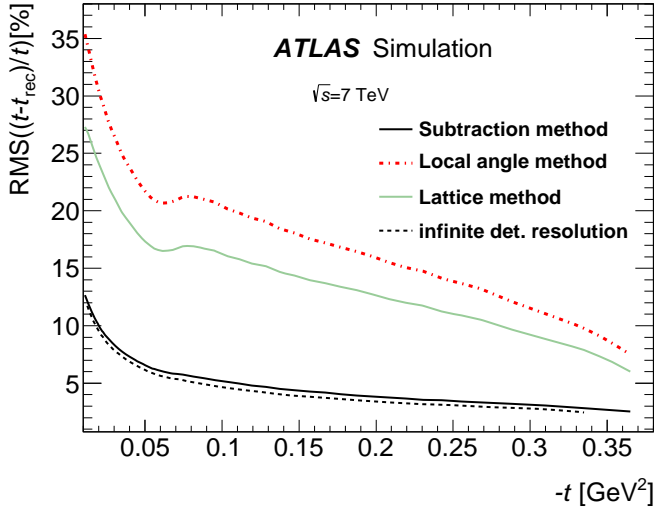}
  \caption{The relative resolution $\mathrm{RMS}((t-t_\mathrm{rec})/t)$, where $t_\mathrm{rec}$ is the reconstructed 
  value of $t$, for different reconstruction methods. The resolution for local 
   subtraction is not shown as it is practically the same as for the lattice method. 
   The dotted line shows the resolution without detector resolution, accounting only for 
   beam effects. It is the same for all methods.} 
  \label{fig:tres}
\end{figure}

The measured $t$-spectrum is affected by detector resolution and beam smearing 
effects, including angular divergence, vertex smearing and energy smearing.
These effects are visible in the $t$-resolution and the purity of the 
$t$-spectrum. The purity is defined as the ratio of the number of events generated and reconstructed  
in a particular bin to the total number of events reconstructed in that bin.  
The purity is about 60\% for the subtraction method, and is about a factor of two worse 
for the local angle method. 
The limited $t$-resolution induces migration effects between bins, which reduces the purity. 
Figure~\ref{fig:tres} shows the $t$-resolution for different $t$-reconstruction methods.

The resolution for the subtraction method improves from $12\%$ at small $|t|$  
to $3\%$ at large $|t|$ and is about a factor three to four better than the other methods. 
The $t$-resolution differences arise because of differences in the spatial and angular resolution of the various
reconstruction methods. The subtraction method is superior to the others because only the 
spatial resolution contributes, whereas the poor angular resolution 
in the horizontal plane degrades the resolution of the other methods.

The measured $t$-spectrum in each arm, after background subtraction, is corrected for migration 
effects using an iterative, dynamically stabilized unfolding method~\cite{IDS}. 
Monte Carlo simulation is used to obtain the migration matrix used in the unfolding. 
The results are cross-checked using an unfolding based on the singular value decomposition method~\cite{SVD}. 
The unfolding procedure is applied to the distribution obtained using all selected events, after background 
subtraction in each elastic arm.

A data-driven closure test is used to evaluate any bias in the unfolded data spectrum shape due 
to mis-modelling of the reconstruction-level spectrum shape in 
the simulation. The simulation is reweighted at particle level such that the 
reconstructed simulation matches the data.  
The modified reconstruction-level simulation is unfolded 
using the original migration matrix, and the result is compared with the modified particle-level spectrum. 
The resulting bias is considered as a systematic uncertainty.
An additional closure test, based on Monte Carlo simulation, was performed with independent 
Monte Carlo samples with a different physics model with different nuclear slopes.

Further systematic uncertainties are related to the simulation  
of the detector resolution and beam conditions, as discussed in Section~\ref{sec:tspect}.
The systematic shifts are smaller than $0.5\%$ in the $|t|$-range below 
0.2~$\GeV^2$ and increase up to $3\%$ at large $|t|$ for all methods of $t$-reconstruction. 

The unfolding method introduces correlations between bins of the $t$-spectrum. 
These correlations are calculated using simulated pseudo-experiments with 
the same number of events as the data. The resulting statistical covariance matrix 
is included in the fits for the total cross section.

\section{Beam optics}
\label{sec:optics}
The precision of the $t$-reconstruction depends on knowledge of the elements of the transport matrix. 
From the design of the 90~m beam optics along with the alignment parameters of the magnets, the magnet currents and the 
field calibrations, all transport matrix elements can be calculated. 
This initial set of matrix elements is referred to as ``design optics''. 
Small corrections, allowed within the range of the systematic uncertainties, 
need to be applied to the design optics for the measurement of $\sigmatot$.
In particular, corrections are needed in 
the horizontal plane where the phase advance is close to $180^\circ$, because the lever arm 
$M_{12}=\sqrt{\beta\times\betastar}\sin{\Psi}$ is rather 
sensitive to the value of $\Psi$. 

Constraints on beam optics parameters 
are derived from the ALFA data, exploiting the fact that the reconstructed scattering angle 
must be the same for different reconstruction methods using different transport matrix elements.   
The beam optics parameters are determined from a global fit, using these constraints, with the design optics as 
a starting value.  

\subsection{ALFA constraints}
\label{sec:constraints}
The reconstructed tracks from elastic collisions can be used 
to derive certain constraints on the beam optics directly from the data. 
Two classes of constraints are distinguished:
\begin{itemize}
\item Correlations between positions or angles measured either at the A-side and C-side 
or at inner and outer stations of ALFA. These are used to infer the ratio of matrix elements in the beam transport matrix. 
The resulting constraints are independent of any optics input.  
\item Correlations between the reconstructed scattering angles. These are calculated using different methods 
to derive further constraints on matrix elements as scaling factors. These factors 
indicate the amount of scaling needed for a given matrix element ratio in order 
to equalize the measurement of the scattering angle. These constraints depend on the given optics model.
The design beam optics with quadrupole currents measured during the run is used as reference to calculate the constraints. 
\end{itemize}  
With parallel-to-point focusing the measured position at the RP is to a first approximation 
related to the scattering angle 
by $w = M_{12,w} \times \theta^{\star}_w \; , \; \; w\in\{x,y\} \;$,  
up to a small contribution from the vertex term in $M_{11}$ in the horizontal plane. 
The ratio of A-side to the C-side track positions is thus on average equal to the ratio of 
the lever arms $M^\mathrm{A}_{12}/M^\mathrm{C}_{12}$, because the scattering angle at the A-side 
is the same as at the C-side for elastic scattering, up to beam divergence effects. 
In a similar way, the ratio of the $M_{22}$ matrix elements is obtained from the correlation between 
the angles measured in the two stations on one side. 
In the vertical plane the ratio of $M_{12}$ between 
the inner and outer station is also measured, whereas in the horizontal plane the correlation between 
positions cannot be translated into a measurement of the $M_{12}$ ratio because of the contribution 
from the vertex term in $M_{11}$, which is different for the inner and outer detectors. 
\begin{figure}[h!]
  \centering
  \includegraphics[width=100mm]{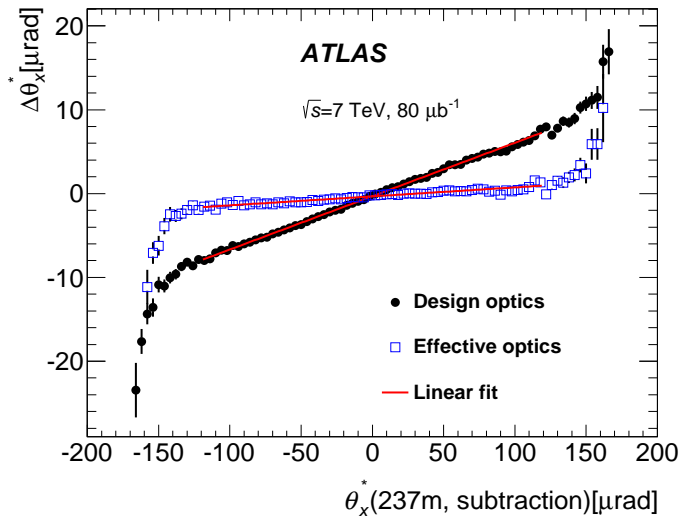}
  \caption{The difference in reconstructed scattering angle $\Delta\theta^\star_x$ between the subtraction and local angle 
  methods as a function 
  of the scattering angle from subtraction method for the inner detectors. In each bin of the scattering angle 
  the points show
  the mean value of $\Delta\theta^\star_x$ 
  and the error bar represents the RMS. The line represents the result of a linear fit. 
  Values obtained using the effective optics (see Section \ref{sec:optics_fit}) are also shown for comparison.} 
  \label{fig:per_x}
\end{figure}

The second class of constraints is derived from the assumption that the reconstructed scattering angle must 
be the same for different methods for a consistent beam optics model. The best example is the comparison
of the scattering angle in the horizontal plane reconstructed with the subtraction method, Eq.~\eqref{eq:subtraction}, 
which is based on the position and $M_{12,x}$,  
and the local angle method Eq.~\eqref{eq:localangle}, which is based on the local angle and $M_{22,x}$. 
The scaling factor for the matrix element ratio $M_{12}/M_{22}$ is derived from the slope 
of the difference of the scattering angle between the two methods as a function of the scattering 
angle determined with the subtraction method, as shown in Fig.~\ref{fig:per_x}.
Here the slope of about $5\%$ indicates that the design optics ratio $M_{12,x}/M_{22,x}$ needs to 
be increased by $5\%$ in order to obtain from the data, on average, the same scattering angle 
from both methods. As discussed in Section~\ref{sec:optics_fit} such an increase in the ratio is 
compatible with realistic deviations of the quadrupole strengths from nominal. 
Constraints of this type are obtained for inner and outer detectors in both the 
vertical and horizontal planes independently. 

Finally, a constraint on the ratio of the $M_{12}$ matrix element in the vertical plane 
to that in the horizontal plane is derived  
from the isotropy of the scattering angle in the transverse plane. Here 2D-patterns of the horizontal 
and vertical scattering angle components reconstructed with the subtraction method are analyzed by selecting regions 
with approximately constant density, which appear as sections of a circle in the case of perfect optics. 
A scaling factor for $M_{12,y}/M_{12,x}$ is inferred from a fit of an ellipse to these patterns, 
and the ratio of matrix elements is taken from the ratio of the major to minor axis of the ellipse. 
All ALFA constraints on the beam optics are summarized in Table~\ref{tab:constraints}. 

\begin{table}
  \begin{center}
    \begin{tabular}{lcccc}
      \hline \hline
Constraint     & Value & Stat. & Syst. & Total \\ 
      \hline
$M_{12,x}(237\mbox{ m}) B_2/B_1$      & 1.0063 & 0.0015 & 0.0041 & 0.0044\\
$M_{12,x}(241\mbox{ m}) B_2/B_1$      & 1.0034 & 0.0010 & 0.0041 & 0.0042\\ 
$M_{22,x} B_2/B_1$                   & 0.9932 & 0.0007 & 0.0041 & 0.0042 \\ 
$M_{12,y}(237\mbox{ m}) B_2/B_1$      & 0.9951 & 0.0001 & 0.0026 & 0.0026 \\
$M_{12,y}(241\mbox{ m}) B_2/B_1$      & 0.9972 & 0.0001 & 0.0026 & 0.0026\\
$M_{12,y} 237/241 B_2$               & 1.0491 & 0.0001 & 0.0007 & 0.0008 \\ 
$M_{12,y} 237/241 B_1$               & 1.0481 & 0.0001 & 0.0007 & 0.0008 \\ 
$M_{22,y} B_2/B_1$                   & 0.9830 & 0.0002 & 0.0180 & 0.0180 \\ \hline
$R(M_{12,x}/M_{22,x}) (237\mbox{ m}) $   & 1.0551 & 0.0003 & 0.0022 & 0.0023\\
$R(M_{12,x}/M_{22,x}) (241\mbox{ m}) $   & 1.0453 & 0.0002 & 0.0013 & 0.0014\\ 
$R(M_{12,y}/M_{22,y}) (237\mbox{ m}) $   & 1.0045 & 0.0001 & 0.0061 & 0.0061\\
$R(M_{12,y}/M_{22,y}) (241\mbox{ m}) $   & 1.0046 & 0.0001 & 0.0065 & 0.0065 \\ 
$R(M_{12,y}/M_{12,x}) (237\mbox{ m}) $   & 0.9736 & 0.0052 & 0.0104 & 0.0116\\ 
$R(M_{12,y}/M_{12,x}) (241\mbox{ m}) $   & 0.9886 & 0.0057 & 0.0072 & 0.0092\\ 
      \hline\hline
    \end{tabular}
  \caption{Summary of the ALFA constraints on beam optics with uncertainties. The first group 
  of constraints are for transport matrix element ratios, the second group comprises scaling factors 
  for matrix element ratios with respect to design optics. $B_1$ and $B_2$ represent beam 1 and beam 2. 
}
  \label{tab:constraints}
  \end{center}
\end{table}

Systematic uncertainties are obtained from several variations of the analysis and 
dominate the precision of the constraints. An important uncertainty is deduced from 
the difference between the constraints for the two arms, i.e.\ the difference between the upper and lower 
detectors for which the optics must be the same. A variation of the selection cuts is used to probe the 
possible influence of background. All constraints are obtained from 
fits and a variation of the fit range
allows potential biases related to acceptance effects to be tested.
All alignment parameters are varied within their systematic uncertainties and the constraint 
analysis repeated. The observed maximum change in each constraint is taken as the corresponding systematic uncertainty. 
The limited detector resolution induces in the scaling factors
a small bias of about $0.5\%$ even with perfect optics. 
This is estimated with simulation and subtracted. 
This correction depends on the physics model and detector resolution used in the simulation. 
The simulation was repeated with a variation of the nuclear slope $B=19.5\pm1.0\GeV^{-2}$ and 
separately with a variation of the detector resolution according to the procedure outlined 
in Section~\ref{sec:sim}. In both cases the maximum change of the constraints with 
the alternative bias corrections is taken as a systematic uncertainty. 
For some of the constraints with a similar type of uncertainty the systematic errors are averaged to suppress 
statistical fluctuations.

\subsection{Beam optics fit}
\label{sec:optics_fit}

The constraints described in the previous section are combined in a fit 
used to determine the beam optics. 
The free parameters of the fit are the quadrupole strengths in both beams.
All ALFA constraints are treated as uncorrelated. 
The effect of the longitudinal quadrupole position is negligible when varied by its 
uncertainty, though this is considered in the total systematic uncertainties (see Section \ref{sec:tsyst}).
In the minimization procedure, the beam 
optics calculation program MadX is used to extract the optics parameters and  
to calculate the matrix element ratios for a given set of magnet strengths.

The ALFA detection system provides precise constraints on the matrix element ratios, 
but cannot probe the deviation of single magnets. Therefore several sets of optics parameters 
exist that minimize the $\chi^2$, arising from different 
combinations of magnet strengths.
The chosen configuration, 
called the effective optics, is one solution among many. This solution is obtained by allowing only the inner 
triplet magnets Q1 and Q3 to vary coherently from their nominal strength. Q1 and Q3 were manufactured at a different site from 
the other quadrupoles, and relative calibration differences are possible. 
Other alternatives 
are taken into account in the total systematic uncertainties (see Section \ref{sec:tsyst}).
\begin{figure}[h!]
  \centering
  \includegraphics[width=\textwidth]{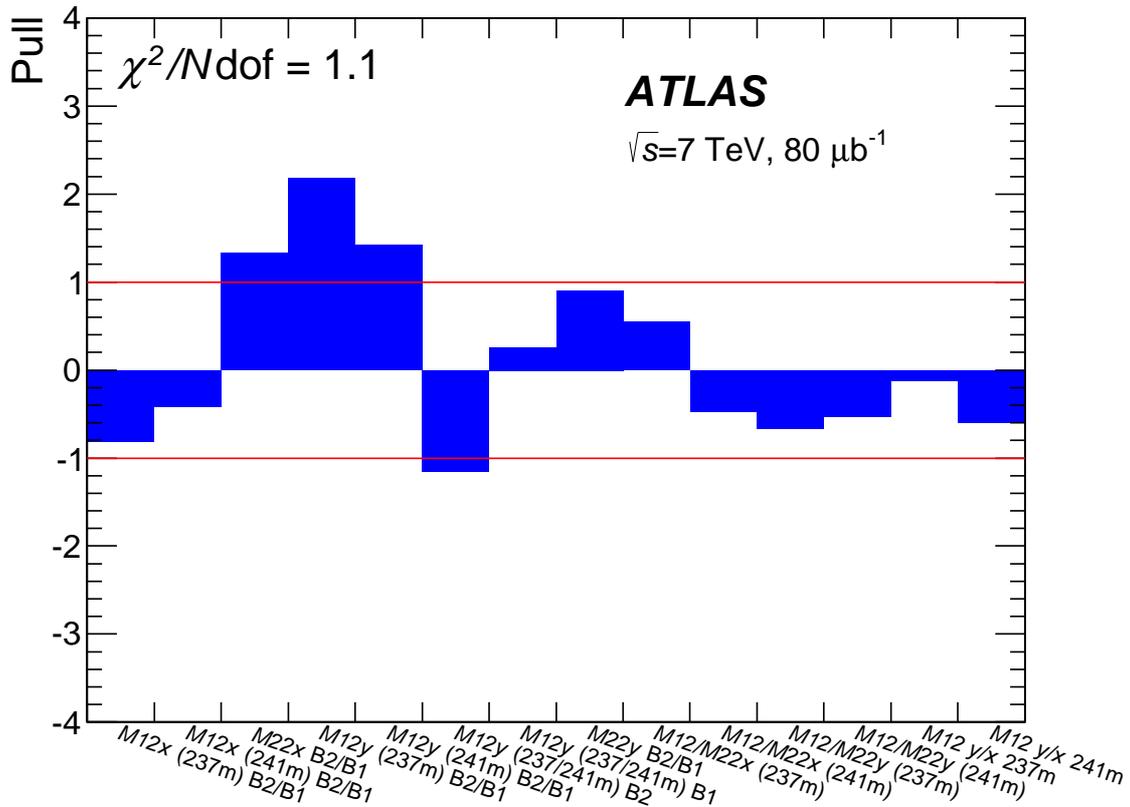}
  \caption{Pulls from the beam optics fit to the ALFA constraints which are given in Table~\ref{tab:constraints}.} 
  \label{fig:pull}
\end{figure}

Figure~\ref{fig:pull} shows the pull of the ALFA constraints after the minimization, which resulted in an 
offset of approximately 0.3\% for the strength of Q1 and Q3, with a difference of about $10\%$ between 
the two beams. 
The $\chi^{2}$ of the fit includes the systematic uncertainties of the constraints and is of good quality 
with $\chi^{2}/N_{\mbox{dof}}=13.2/12$. 
No measurement deviates from the fit by more than about two standard deviations; the largest 
deviation is observed for the high-precision constraint on $M_{12}$ in the vertical plane.  
This solution was cross-checked using the LHC measurements of the phase advance.
This effective optics is used for the total cross-section measurement.

\section{Luminosity determination}
\label{sec:luminosity}

In normal running conditions at high luminosity ($L\, >  10^{33}$\,cm$^{-2}$\,s$^{-1}$), ATLAS
exploits several detectors and algorithms to determine the luminosity and evaluate the related systematic uncertainty.
These include LUCID (luminosity measurement with a Cherenkov integrating detector),  
BCM (beam conditions monitor) and the inner
detector, for the bunch-by-bunch luminosity determination, and the tile and forward
calorimeters, for bunch-integrated luminosity measurements. Details of the
ATLAS luminosity measurement can be found in Ref.~\cite{LumiPaper2011}, which includes a description of all the detectors and
algorithms, the calibration procedure, the background evaluation and subtraction and the estimation of the systematic 
uncertainties.

\par
The conditions in the low-luminosity run analyzed here are very different from those in high-luminosity runs. 
The instantaneous luminosity is about six orders of magnitude lower ($L\,\sim 5 \times 10^{27}$\,cm$^{-2}$\,s$^{-1}$) 
which makes the calorimeter methods unusable due to the lack of sensitivity. 
An additional method based on vertex counting 
in the inner detector (ID) was included; this method is most effective at low pile-up. 
Another difference with respect to the normal high-luminosity conditions is the background composition: 
the beam--gas contribution, normally negligible, can become competitive with the collision rate in the low-luminosity 
regime. The background due to slowly decaying, collision-induced radiation 
(often called ``afterglow'' ~\cite{LumiPaper2011}) becomes conversely less important, because of 
the presence of only a few colliding bunches. 

\par
In the 2011 data taking, the BCM was used as the baseline detector for the luminosity determination. It consists of four independent 
detectors grouped into two sets of two, vertical (BCMV) 
and horizontal (BCMH), located on each side of the interaction point and made of diamond sensors. 
For the high-\betastar\ run, an inclusive-OR of the two sides was used to define an event with activity in the detector. 
The luminosity is determined with the event-counting method based on this definition (BCMV\_EventOR, BCMH\_EventOR). 
LUCID is also located on both sides of the interaction point and detects charged particles produced 
in the forward direction by collecting, with photomultipliers, the Cherenkov light produced. 
It measures luminosity with the same definition as used for BCM (LUCID\_EventOR), with the addition 
of an event-counting algorithm, requiring a coincidence between the two sides (LUCID\_EventAND), and a
hit-counting algorithm (LUCID\_HitOR), in which the number of photomultipliers providing a signal above threshold is counted.
A third method for measuring the per-bunch luminosity is provided by the inner detector. This method counts the number of primary 
vertices per event, which is proportional to the luminosity. The vertex selection criteria required a minimum of five good-quality tracks with 
transverse momentum larger than 400 \MeV, forming a common vertex~\cite{LumiPaper2011}. 
As about 12\% of the  high-\betastar\ run data were acquired when the high voltage of the ID was lowered for detector-protection reasons, 
the vertex-based algorithm is not available for that part of the run.

\par
\begin{figure}[h!]
  \centering
  \includegraphics[width=100mm]{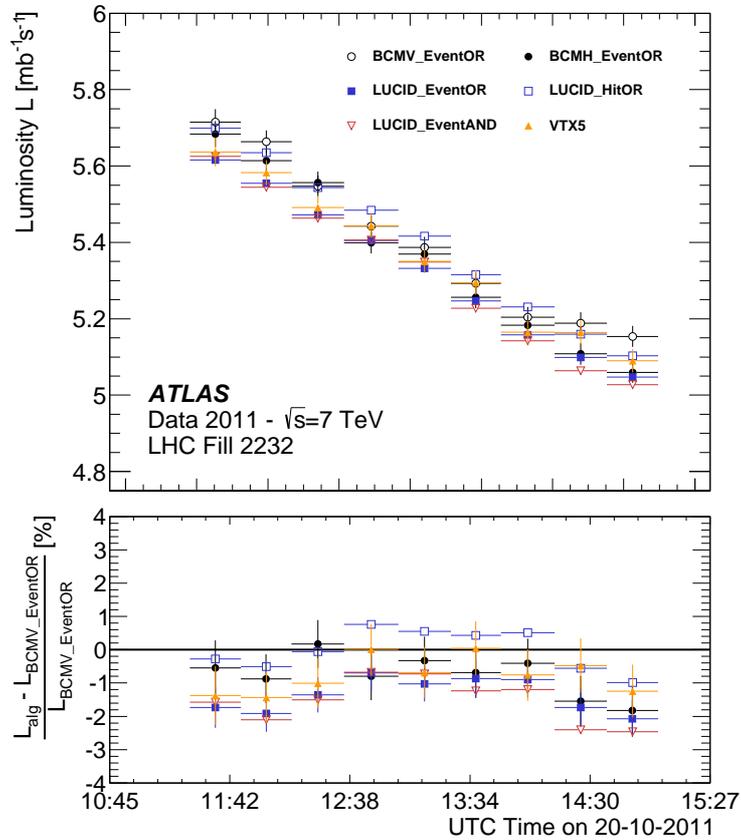}
\caption{Luminosity measured by the various algorithms (top) and relative deviations from the reference BCMV\_EventOR 
algorithm (bottom), as a function of time.}
\label{fig:LumiVsLB}
\end{figure}

The absolute luminosity scale of each algorithm was calibrated~\cite{LumiPaper2011} by the van der Meer ({\it vdM}) method 
in an intermediate luminosity regime ($L\,\sim 5 \times 10^{30}$\,cm$^{-2}$\,s$^{-1}$). 
Table~\ref{tab:LumiTable} lists the integrated luminosity reported by the BCM, LUCID and vertex-based luminosity algorithms 
(VTX5) during the run analyzed here, 
both for the full sample and for the fraction during which the ID was fully operational. 
As for the standard running conditions, BCMV\_EventOR was chosen as the preferred algorithm for the luminosity
determination, as its response is stable and independent of the luminosity scale, from {\it vdM} to high luminosity~\cite{LumiPaper2011} conditions. 
Table~\ref{tab:LumiTable} also reports the fractional deviation of each measurement from the reference value; 
the largest such difference is $1.6\%$. In Fig.~\ref{fig:LumiVsLB} the luminosity measurements from the various algorithms are 
shown as a function of time (top), together with the percent deviations from the reference algorithm (bottom).

\begin{table}[htbp]
\begin{center}
\begin{tabular}{ l | c|c  |c|c }
\hline    
\hline
    Algorithm & $L_{\mathrm{int}}$ [$\mu$b$^{-1}$]  & Deviation ($\%$) & $L_{\mathrm{int}}$ [$\mu$b$^{-1}$]  & Deviation ($\%$)  \\ 
\hline
                    &  \multicolumn{2}{c|}{Full data set} & \multicolumn{2}{c}{Data with ID on}    \\ \hline
    BCMV\_EventOR     & $78.72 \pm 0.13$  & -       & $69.48 \pm 0.12$ &  - \\
    BCMH\_EventOR     & $78.10 \pm 0.13$  & $-0.80$ & $68.94 \pm 0.12$ &  $-0.78$ \\
    LUCID\_EventOR    & $77.63 \pm 0.04$  & $-1.38$ & $68.55 \pm 0.04$ &  $-1.34$ \\ 
    LUCID\_EventAND   & $77.49$           & $-1.57$ & $68.42$          &  $-1.53$ \\
    LUCID\_HitOR      & $78.71 \pm 0.03$  & $-0.01$ & $69.49 \pm 0.03$ &  $+0.01$  \\
    VTX5                           & &              & $68.93 \pm 0.13$ & $-0.79$ \\
    \hline
    \hline
\end{tabular}
\end{center} 
\caption{Integrated luminosity measured using the different algorithms and deviations from the reference value (BCMV\_EventOR). 
The results are reported for the full data set and for the subset thereof during which the ATLAS ID was fully operational. 
The uncertainties are statistical only.
For LUCID\_EventAND, no error is assigned as the luminosity is numerically determined from a 
formula dependent on the measured event rate.}
\label{tab:LumiTable}
\end{table} 

The contributions to the systematic uncertainty affecting the absolute integrated luminosity during the high-\betastar\ 
run can be categorized as follows.
\begin{itemize}
\item
The uncertainty on the absolute luminosity scale, as determined by the {\it vdM} method, amounts to 1.53\%~\cite{LumiPaper2011}. 
Because it is dominated by beam conditions rather than by instrumental effects, this ``scale uncertainty'' is common to all luminosity algorithms.
\item
The ``calibration-transfer'' uncertainty, associated with transferring the absolute luminosity scale from the intermediate-luminosity regime of 
the {\it vdM} scans to the very low-luminosity conditions of elastic-scattering measurements five months later. 
It is discussed more extensively below.
\item
The uncertainty related to the subtraction of beam-associated background during the high-\betastar\ run is
estimated to be 0.20\%, which results from varying the magnitude of the background correction for the reference 
algorithm by 80\%.
\end{itemize}
The calibration-transfer uncertainty reflects the uncertainty in the potential shift in BCM response from {\it vdM} to high-luminosity 
conditions (0.25\%), 
the relative long-term stability (0.70\%) of the BCM during the several months of high-luminosity running that separate 
the {\it vdM} calibration period from the high-\betastar\ run, and the stability of the LUCID and BCM calibration transfer 
from the high- to the very low-luminosity regime. While the former two are extensively documented in Ref.~\cite{LumiPaper2011}, 
the latter is more difficult to assess. The consistency between independent estimates of the integrated luminosity, 
as quantified in Table~\ref{tab:LumiTable} by the largest deviation from the reference value (1.6\%),  is therefore 
conservatively taken as an upper limit on the systematic uncertainty associated with the third step (high to very-low luminosity) of this calibration transfer. 

\par
The total systematic uncertainty on the integrated luminosity $L_{\mathrm{int}}$ during the high-\betastar\ run is computed 
as the sum in quadrature  
of the scale uncertainty, the overall calibration-transfer uncertainty and the background uncertainty; it amounts to 2.3\%. 
The final result for the selected running period is:
\begin{equation*}
L_{\mathrm{int}}=78.7 \pm 0.1 \,(\mathrm{stat.}) \pm 1.9 \,(\mathrm{syst.})  ~\mu \mathrm{b}^{-1} \; .\label{LumiWithSys}
\end{equation*}

\section{The differential elastic cross section}
\label{sec:tspect}
\begin{figure}[h!]
  \centering
  \includegraphics[width=\textwidth]{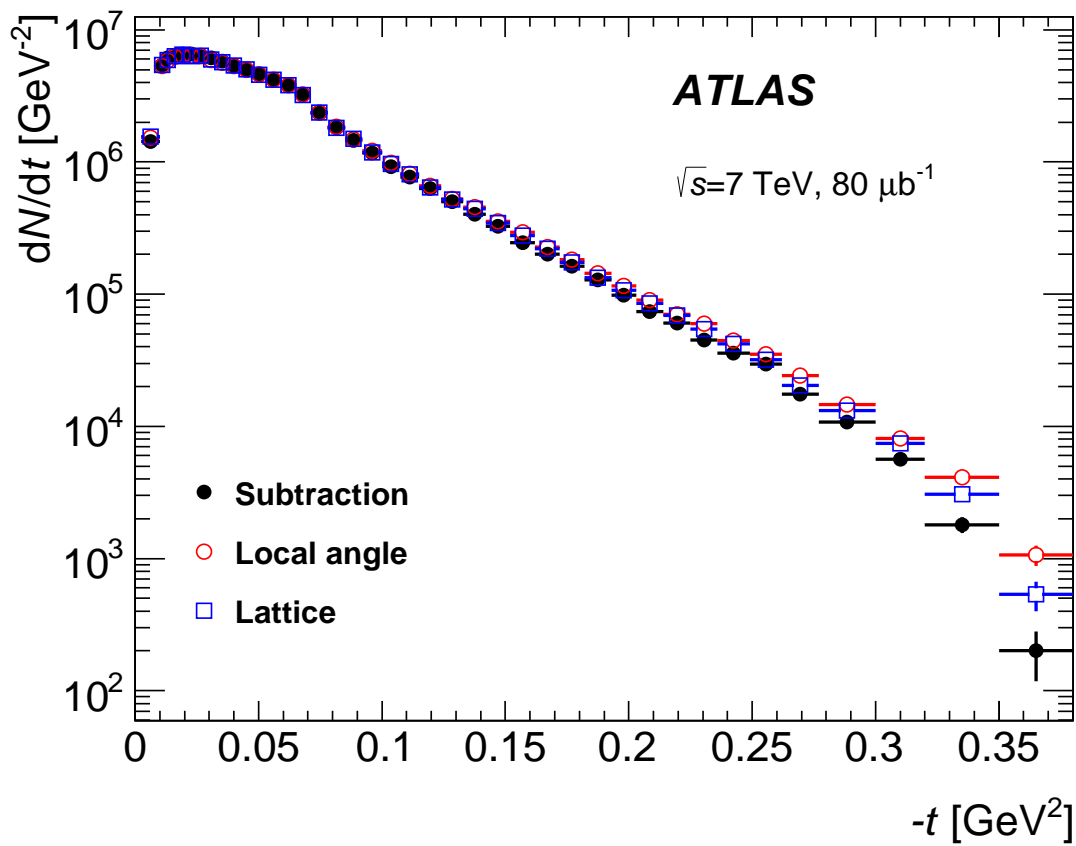} 
  \caption{Counting rate $\mathrm{d}N/\mathrm{d}t$ as a function of $t$ 
in arm 1 for different reconstruction methods before corrections. 
  The error bars represent the statistical uncertainty only. 
  The local subtraction method is not shown as it is indistinguishable from the lattice method.} 
  \label{fig:traw_arm1}
\end{figure}

The raw $t$-spectrum of elastic-scattering candidates in one detector arm after the event selection is shown in 
Fig.~\ref{fig:traw_arm1}. Different $t$-reconstruction methods using the effective optics are compared. 
Differences between the methods are visible at large $|t|$, where the resolution effects for methods using the local angle 
are most important. 
In order to calculate the differential elastic cross section, several corrections are applied. 
The corrections are done individually per detector arm and the corrected spectra from the two arms are combined. 
In a given bin $t_i$ the cross section is calculated according to the following formula:
\begin{equation}\label{eq:cross-section}
\frac{\mathrm{d}\sigma}{\mathrm{d}t_i} = \frac{1}{\Delta t_i}\times \frac{{\cal M}^{-1}[N_i - B_i]}{A_i  \times \epsilon^{\mathrm{reco}} \times \epsilon^{\mathrm{trig}} \times \epsilon^{\mathrm{DAQ}}  \times L_{\mathrm{int}} }\; \; , 
\end{equation}
where $\Delta t_i$ is the bin width, ${\cal M}^{-1}$ represents the unfolding procedure applied to the 
background-subtracted number of events $N_i - B_i$, $A_i$ is the acceptance,  
$\epsilon^{\mathrm{reco}}$ is the event reconstruction efficiency, $\epsilon^{\mathrm{trig}}$ is the trigger efficiency, 
$\epsilon^{\mathrm{DAQ}}$ is the dead-time correction and $L_{\mathrm{int}}$ is the integrated luminosity used for this
analysis. 

\begin{figure}[h!]
  \centering
  \includegraphics[width=0.8\textwidth]{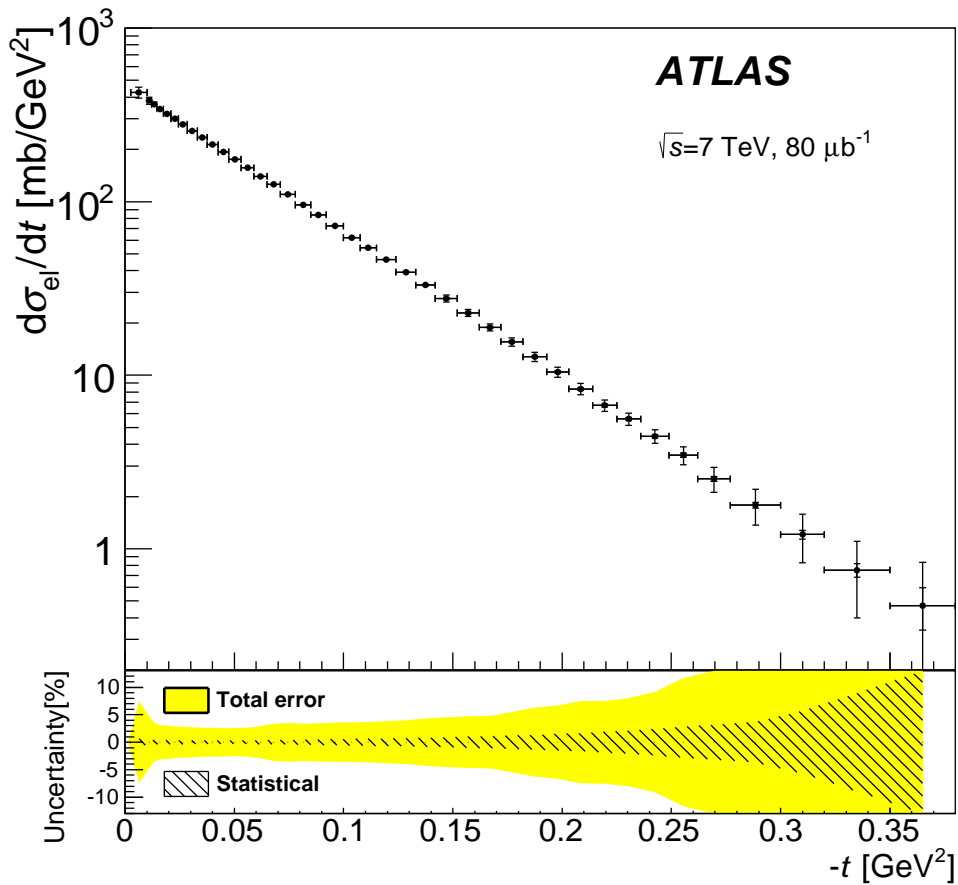}
  \caption{The differential elastic cross section measured using the subtraction method. 
   The outer error bars indicate the total experimental uncertainty and the inner error bars the 
   statistical uncertainty. The lower panel shows the relative total and statistical uncertainties.} 
  \label{fig:differential_elastic_crosssection}
\end{figure}

The binning in $t$ is appropriate for the experimental resolution and statistics. At small $t$ the selected bin width is 1.5 times the resolution.
At larger 
$|t|$ the bin width is increased to compensate for the lower number of events from the exponentially falling distribution. 
The resulting differential elastic cross section using the subtraction method is shown in 
Fig.~\ref{fig:differential_elastic_crosssection} with statistical and systematic uncertainties.
The numerical values for all bins are summarized in Table~\ref{tab:differential_elastic_crosssection}. 

\begin{figure}[h!]
  \centering
  \includegraphics[width=12cm]{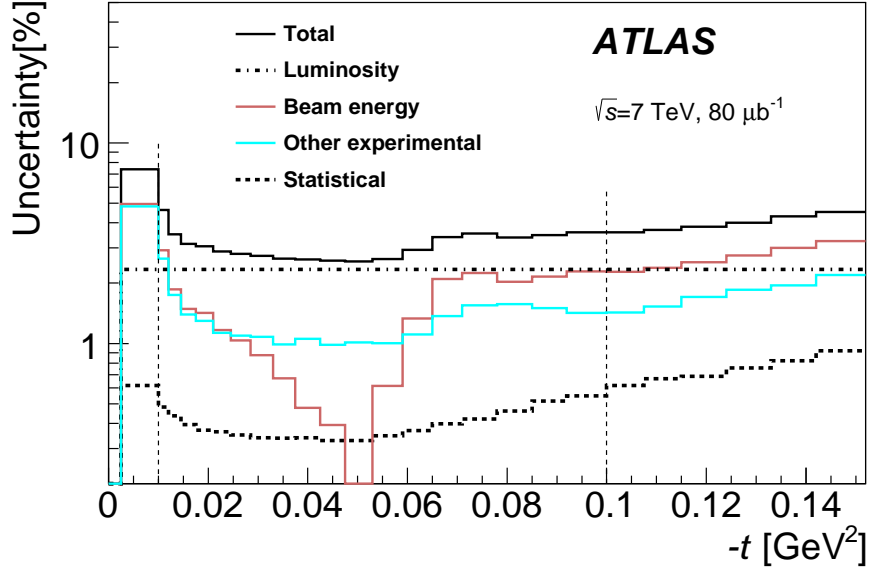}
  \caption{Relative statistical and systematic uncertainties for the differential elastic cross section 
   for different groups of uncertainties. Only the $t$-range relevant to the extraction of the total cross section is shown; 
the vertical lines indicate the fit range.} 
  \label{fig:tsyst_rel_uncertainty}
\end{figure}

An overview of the contributions to the total uncertainty is given in Fig.~\ref{fig:tsyst_rel_uncertainty}. 
Inside the fit range used to extract the total cross section, indicated by vertical lines, the dominant 
contribution to the uncertainty is the $t$-independent luminosity error. 
The statistical error is only a small contribution to the total error. 
The experimental uncertainty comprises various contributions which are discussed in the following. 
\begin{table}[h!]
  \begin{center}
\hspace*{-0.1cm}  
    \begin{tabular}{ccccccc}
      \hline \hline
Low $|t|$ edge  & High $|t|$ edge & Central $|t|$ & $\mathrm{d}\sigma_{\mathrm{el}}/\mathrm{d}t$ & Stat. uncert.  & Syst. uncert. & Total uncert.\\
$[$\GeV$^2$]   & [\GeV$^{2}$] &  [\GeV$^{2}$] & [mb/\GeV$^2$] & [mb/\GeV$^2$] & [mb/\GeV$^2$] & [mb/\GeV$^2$] \\ \hline
0.0025 & 0.0100 & 0.0062 & 425.8 & 2.6 & 31.3 & 31.4 \\
0.0100 & 0.0120 & 0.0110 & 382.6 & 1.9 & 17.7 & 17.8 \\
0.0120 & 0.0145 & 0.0132 & 363.2 & 1.6 & 12.6 & 12.7 \\
0.0145 & 0.0175 & 0.0160 & 342.0 & 1.4 & 10.6 & 10.7 \\
0.0175 & 0.0210 & 0.0192 & 320.9 & 1.2 & 9.7 & 9.8 \\
0.0210 & 0.0245 & 0.0227 & 300.3 & 1.1 & 8.6 & 8.6 \\
0.0245 & 0.0285 & 0.0265 & 279.4 & 1.0 & 7.8 & 7.8 \\
0.0285 & 0.0330 & 0.0307 & 256.1 & 0.9 & 7.0 & 7.0 \\
0.0330 & 0.0375 & 0.0352 & 234.3 & 0.8 & 6.2 & 6.2 \\
0.0375 & 0.0425 & 0.0400 & 213.2 & 0.7 & 5.5 & 5.6 \\
0.0425 & 0.0475 & 0.0450 & 193.3 & 0.6 & 5.0 & 5.0 \\
0.0475 & 0.0530 & 0.0502 & 175.1 & 0.6 & 4.5 & 4.5 \\
0.0530 & 0.0590 & 0.0559 & 156.5 & 0.5 & 4.1 & 4.1 \\
0.0590 & 0.0650 & 0.0619 & 139.8 & 0.5 & 4.1 & 4.1 \\
0.0650 & 0.0710 & 0.0679 & 125.5 & 0.5 & 4.2 & 4.3 \\
0.0710 & 0.0780 & 0.0744 & 110.2 & 0.5 & 3.9 & 3.9 \\
0.0780 & 0.0850 & 0.0814 & 95.8 & 0.4 & 3.2 & 3.2 \\
0.0850 & 0.0920 & 0.0884 & 83.9 & 0.4 & 2.9 & 2.9 \\
0.0920 & 0.1000 & 0.0959 & 72.4 & 0.4 & 2.6 & 2.6 \\
0.1000 & 0.1075 & 0.1037 & 62.0 & 0.4 & 2.2 & 2.2 \\
0.1075 & 0.1150 & 0.1112 & 54.1 & 0.4 & 2.0 & 2.0 \\
0.1150 & 0.1240 & 0.1194 & 46.4 & 0.3 & 1.8 & 1.8 \\
0.1240 & 0.1330 & 0.1284 & 39.2 & 0.3 & 1.5 & 1.6 \\
0.1330 & 0.1420 & 0.1374 & 33.0 & 0.3 & 1.4 & 1.4 \\
0.1420 & 0.1520 & 0.1468 & 27.7 & 0.3 & 1.2 & 1.3 \\
0.1520 & 0.1620 & 0.1568 & 22.8 & 0.2 & 1.1 & 1.1 \\
0.1620 & 0.1720 & 0.1668 & 18.88 & 0.21 & 0.88 & 0.91 \\
0.1720 & 0.1820 & 0.1768 & 15.58 & 0.19 & 0.83 & 0.85 \\
0.1820 & 0.1930 & 0.1873 & 12.77 & 0.17 & 0.79 & 0.80 \\
0.1930 & 0.2030 & 0.1978 & 10.45 & 0.16 & 0.68 & 0.70 \\
0.2030 & 0.2140 & 0.2083 & 8.33 & 0.14 & 0.61 & 0.62 \\
0.2140 & 0.2250 & 0.2193 & 6.70 & 0.13 & 0.49 & 0.50 \\
0.2250 & 0.2360 & 0.2303 & 5.60 & 0.12 & 0.43 & 0.45 \\
0.2360 & 0.2490 & 0.2422 & 4.45 & 0.11 & 0.39 & 0.41 \\
0.2490 & 0.2620 & 0.2553 & 3.46 & 0.10 & 0.39 & 0.40 \\
0.2620 & 0.2770 & 0.2691 & 2.53 & 0.08 & 0.41 & 0.41 \\
0.2770 & 0.3000 & 0.2877 & 1.78 & 0.07 & 0.41 & 0.42 \\
0.3000 & 0.3200 & 0.3094 & 1.21 & 0.07 & 0.37 & 0.38 \\
0.3200 & 0.3500 & 0.3335 & 0.75 & 0.07 & 0.35 & 0.35 \\
0.3500 & 0.3800 & 0.3636 & 0.47 & 0.13 & 0.34 & 0.36 \\ \hline\hline
    \end{tabular}
  \caption{The measured values of the differential elastic cross section with statistical and systematic uncertainties. 
The central $t$-values in each bin are calculated from simulation, in which a slope parameter of $B = 19.5  \GeV^{-2}$ is used.}
  \label{tab:differential_elastic_crosssection} 
  \end{center}
\end{table}
\subsection{Systematic uncertainties}
\label{sec:tsyst}
The following uncertainties are propagated to the differential elastic cross section:
\begin{itemize}
\item The amount of background is varied by the difference between the yields obtained from 
the anti-golden and the vertex methods. The background shape is varied by the inversion of the sign 
of the vertical coordinate $y$ on one side of the  
anti-golden arm, combined with an inversion of the $x$-coordinate (see Section~\ref{sec:event_selection}). 
\item The impact of alignment uncertainties is estimated from different sets of the alignment parameter values (see Section~\ref{sec:alignment}).  
\item The uncertainties related to the effective optics are obtained by varying the optics constraints, 
changing the strength of the quadrupoles Q2, Q4, Q5 and Q6 by $\pm 1 \permil $
and changing only Q5 and Q6 by $-2\permil $, as suggested by LHC constraints on the phase advance. 
Additional uncertainties are determined by varying  
the quadrupole alignment according to its uncertainty, propagating the fit 
uncertainties for the strength of Q1 and Q3 to the resulting optics and by taking the difference between the 
transfer-matrix-based beam transport used in the optics fit and the MadX beam transport (see Section~\ref{sec:optics}).
\item The nuclear slope used in the simulation is varied conservatively by $\pm1$ \GeV$^{-2}$ around 19.5 \GeV$^{-2}$ 
(see Section~\ref{sec:mc_th}).  
\item The detector resolution is varied, replacing the resolution determined from collision data with values 
3--4 $\mu$m smaller, as predicted from the GEANT4 simulation, and 4--5 $\mu$m larger, as 
measured using test-beam data. Additionally, a $y$-dependent 
resolution is used instead of a constant value (see Section~\ref{sec:sim}).  
\item The emittance used to calculate the angular divergence in the simulation is varied by $\pm 10\%$ (see Section~\ref{sec:beam_conditions}). 
\item The event reconstruction efficiency is varied by its uncertainty discussed in Section~\ref{sec:reco_efficiency}.
\item For the tracking efficiency the minimum number of fibre layers required to reconstruct a track is varied between three and six (see Section~\ref{sec:track_reco}).
\item The intrinsic unfolding uncertainty is determined from the data-driven closure test in Section~\ref{sec:unfolding_acceptance}. 
\item The impact of a residual beam crossing angle in the horizontal plane of $\pm10\, \mu$rad is taken into account. 
This variation is derived from the precision of the beam position monitors.
\item The nominal beam energy used in the $t$-reconstruction according to Eq.~\eqref{eq:t-basic} is changed 
by $0.65\%$ \cite{Wenninger}. 
\item The luminosity uncertainty of 2.3$\%$ is propagated to the cross section (see Section~\ref{sec:luminosity}).
\end{itemize}

\begin{figure}[h!]
  \centering
  \includegraphics[width=12cm]{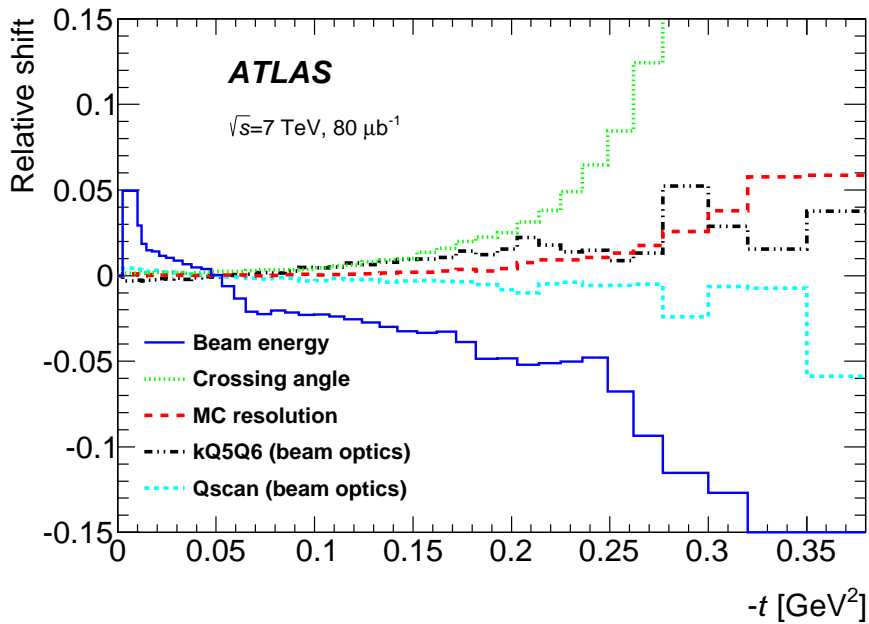}
  \caption{Relative systematic shifts in the differential elastic cross section as a function of $t$ for 
  selected uncertainty sources. Shown are the uncertainties related to the beam optics (kQ5Q6 and Qscan), to the 
  modelling of the detector resolution in the
  simulation (MC resolution), to the crossing angle, and to the beam energy. Here kQ5Q6 refers to changing the 
strength of Q5 and Q6 by $-2\permil$ and Qscan refers to changing the strength of Q2, Q4, Q5, Q6 by $\pm1\permil$.} 
  \label{fig:shifts_part0}
\end{figure}

The systematic uncertainties related to the sources above are calculated by the
offset method. In this method, the nominal value of a certain parameter in the analysis chain is varied according to the 
assigned uncertainty. The shift in bin $i$ for systematic uncertainty source $k$, 
$\delta_k(i)=\mathrm{d}\sigma_k(i)/\mathrm{d}t - \mathrm{d}\sigma_{\mathrm{nominal}}(i)/\mathrm{d}t$ 
 is recorded, keeping track of the sign and thereby accounting 
for correlations across the $t$-spectrum. In total 24 systematic shifts are considered and 
numerical tables with the shifts can be found in HepData~\cite{HepData}.  
Systematic shifts are included in the fit used to determine the total cross section, as outlined in Section~\ref{sec:sigma_tot}. 
The most important $t$-dependent shifts are shown in Fig.~\ref{fig:shifts_part0} for the entire range of the $t$-spectrum; 
the range used to fit $\sigmatot$ is from from $t = -0.01 \GeV^2$ to $t = -0.1 \GeV^2$. 

\section{The total cross section}
\label{sec:sigma_tot}
The total cross section and the slope parameter $B$ are obtained from a fit of the theoretical spectrum 
(Eq. \eqref{eq:tgen}),
including the Coulomb--nuclear interference term, to the measured differential cross section.
The electric form factor $G(t)$, 
the Coulomb phase $\phi(t)$ and the $\rho$-parameter are fixed to the nominal values
as discussed in Section~\ref{sec:mc_th}.
Both the statistical and systematic uncertainties as well as their correlations are taken into account in the fit.
The statistical correlations are included in the covariance matrix calculated in 
the unfolding procedure. The correlations of systematic uncertainties are taken into account by using a 
profile minimization procedure~\cite{profile}, where nuisance parameters corresponding to all 24 systematic shifts are included, 
and the $\chi^2$ is given by:
\begin{eqnarray}\label{eq:profile_fit}
 \chi^2 & = & \sum_{i,j}\left[\left(D(i)-\left(1+\sum_{l=1}^{2} \alpha_l\right) \times T(i) - \sum_{k=1}^{22} \beta_k \times \delta_k(i)\right) \times V^{-1}(i,j)\right. \\ \nonumber
        & &  \left.\times \left(D(j)-\left(1+\sum_{l=1}^{2} \alpha_l\right) \times T(j) - \sum_{k=1}^{22} \beta_k \times \delta_k(j)\right)\right] 
 +\sum_{k=1}^{22}\beta_k^2 + \sum_{l=1}^{2}\frac{\alpha_l^2}{\epsilon_l^2}\; ,  
\end{eqnarray}
where $D(i)$ is the measured value of the elastic cross section in bin $i$, $T(i)$ the theoretical 
prediction and $V(i,j)$ the statistical covariance matrix. For each systematic uncertainty which changes the shape of the $t$-spectrum, a nuisance 
parameter $\beta_k$ multiplying the corresponding shift $\delta_k$ is fitted as a free parameter and 
a penalty term $\sum_k\beta_k^2$ is added to the $\chi^2$. 
Two scale parameters, $\alpha_l$, are used to describe the rescaling of the
normalization of the theoretical prediction due to $t$-independent uncertainties in the luminosity and 
the reconstruction efficiency. The sum in quadrature of these two scale factors divided by their uncertainties 
results in a second penalty term, $\sum_l(\alpha_l^2/\epsilon_l^2)$. Some nuisance parameters in the fit comprise 
a group of parameter variations. For example, 
256 variations of ALFA data constraints by one standard deviation and 256 quadrupole field variations by $\pm 1 \permil$ are 
performed as part of the beam optics uncertainty estimate and are each  
merged into a single nuisance parameter.  

\begin{figure}[h!]
  \centering
  \includegraphics[width=0.9\textwidth]{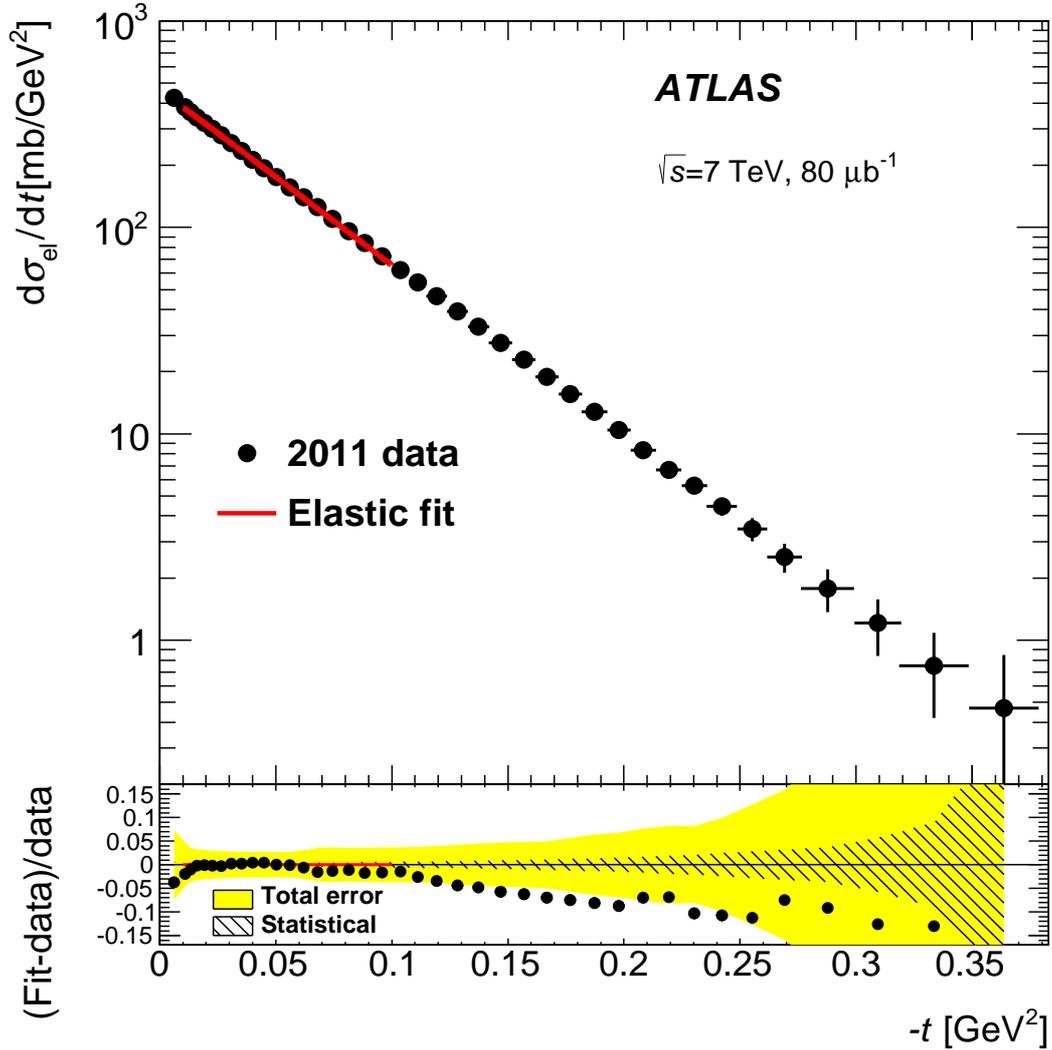}
  \caption{A fit of the theoretical prediction with $\sigmatot$ and $B$ as free parameters to the
 differential elastic cross section reconstructed with the subtraction method.
In the lower panel the points represent the normalized difference between fit and data, 
 the yellow area represents the total experimental uncertainty and the hatched area the statistical component. 
 The red line indicates the fit range, the fit result is extrapolated in the lower panel outside the fit range.} 
  \label{fig:fit_subtraction}
\end{figure}
The differential cross section and the fitted theoretical prediction are shown in
Fig.~\ref{fig:fit_subtraction}.
The fit range is chosen from $t = -0.01 \GeV^2$ to $t = -0.1 \GeV^2$. 
The lower $t$ value is chosen to be as close as possible 
to $t=0$ to reduce the extrapolation uncertainty while having an acceptance above 10\%.
The choice of the upper limit is motivated by theoretical considerations not to extend the fit into 
the region where deviations from the single exponential function are expected
\cite{per_theory}.
The fit yields:
\begin{displaymath}
\sigmatot = 95.35\pm1.30~\mbox{mb} \; \; , \hspace{10mm} B = 19.73\pm0.24 \GeV^{-2} \; ,
\end{displaymath}
where the errors include all statistical and experimental systematic contributions. 
Systematic uncertainties associated with theoretical parameters and the extrapolation $|t|\rightarrow 0$ are 
discussed below. 
The best fit $\chi^2$ is found to be 7.4 for 16 degrees of freedom.  
Important contributions to the $\chi^2$ are related to the alignment, the beam optics and the nominal 
beam energy. 
The results obtained for all $t$-reconstruction methods are compiled in 
Table~\ref{tab:sigma_tot_syst_summary} for $\sigmatot$ and in Table~\ref{tab:B_slope_syst_summary} for $B$. 
\begin{table}
  \begin{center}
    \begin{tabular}{lcccc}
      \hline \hline
            & \multicolumn{4}{c}{$\sigmatot$ [mb]} \\          
	   & Subtraction & Local angle & Lattice & Local subtraction \\ \hline 
Total cross section & 95.35 & 95.57 & 95.03 & 94.98 \\ 
Statistical error & 0.38 & 0.38 & 0.33 & 0.31 \\ 
Experimental error  & 1.25 & 1.36 & 1.30 & 1.30 \\ 
Extrapolation error  & 0.37 & 0.27 & 0.50 & 0.33 \\ \hline
Total  error & 1.36 & 1.44 & 1.43 & 1.38 \\ \hline  \hline
    \end{tabular}
  \caption{The total cross section and uncertainties for four different 
   $t$-reconstruction methods.}
  \label{tab:sigma_tot_syst_summary}
  \end{center}
\end{table}
\begin{table}[h!]
  \begin{center}
    \begin{tabular}{lcccc}
      \hline \hline
            & \multicolumn{4}{c}{$B \; [\GeV^{-2}$]} \\          
           & Subtraction & Local angle & Lattice & Local subtraction \\ \hline 
Nuclear slope & 19.73 & 19.67 & 19.48 & 19.48 \\ 
Statistical error & 0.14 & 0.15 & 0.14 & 0.15 \\ 
Experimental error & 0.19 & 0.26 & 0.22 & 0.21 \\  
Extrapolation error & 0.17 & 0.17 & 0.26 & 0.18 \\ \hline 
Total error & 0.29 & 0.35 & 0.37 & 0.31 \\ \hline  \hline
\end{tabular}
  \caption{The nuclear slope and uncertainties for four different $t$-reconstruction methods.}
  \label{tab:B_slope_syst_summary}
  \end{center}
\end{table}
\subsection{Systematic uncertainties}\label{sec:sigma_tot_syst}
The fit to the $t$-spectrum accounts for the statistical errors in data and Monte Carlo simulation combined with  
24 systematic shifts. 
The statistical component 
of the errors given in Table~\ref{tab:sigma_tot_syst_summary} for the total cross 
section and in Table~\ref{tab:B_slope_syst_summary} for $B$ is estimated using pseudo-experiments. 
The main contributions to the experimental systematic error for the total cross section are the luminosity, nominal 
beam energy and reconstruction efficiency uncertainties. For $B$ the experimental systematic uncertainty 
is dominated by the beam energy uncertainty, whereas other experimental 
errors such as the luminosity uncertainty change only the normalization and have no impact on the slope. 

Additional uncertainties arise from the extrapolation $|t|\rightarrow 0$. 
These are estimated from a variation of the upper end of the fit range 
from $-t=0.1 \GeV^2$ to $-t=0.15 \GeV^2$. 
The upper end at  $-t=0.15 \GeV^2$ is chosen in order to remain in a range where a simple 
exponential is still a reasonable assumption~\cite{per_theory}. 
The upper fit-range edge is also decreased by the same number of bins (six) to $|t|=0.059 \GeV^2$ and the symmetrized 
change is adopted as a systematic uncertainty of approximately 0.4 mb for $\sigmatot$ and 0.17 $\GeV^{-2}$ for $B$. 
The fit-range dependence is the dominant source of uncertainty in the extrapolation.  
As a stability 
check, the lower end of the fit range is varied, leading to a smaller 
change of the total cross section, which is not included in the systematic error. 
Other extrapolation errors are determined from:
the variation of the $\rho$-parameter by $\pm 0.008$, the replacement of the dipole by a double dipole 
parameterization 
for the 
electric form factor and the replacement of the Coulomb phase from West and Yennie \cite{WestAndYennie}
by parameterizations from Refs.~\cite{Cahn,KFK}. 
All these contributions are small compared to the fit-range
variation. The total extrapolation uncertainty is about 0.4 mb for $\sigmatot$ and 0.2 \GeV$^{-2}$ for $B$. 
The total errors given in Table~\ref{tab:sigma_tot_syst_summary} for all $t$-reconstruction methods 
are very similar and amount to $1.5\%$ of the cross-section value. 

Several stability tests were carried out to substantiate the nominal result and its associated 
uncertainty.
\begin{itemize}
\item A fit with only the statistical covariance matrix but ignoring all nuisance parameter contributions was
  performed. The resulting values are given in Table~\ref{tab:fit_stat}. Because the fit ignores 
  systematic errors the $\chi^2$ is larger.
\item The independent samples in each detector arm were corrected and analyzed independently. 
\item An alternative procedure was performed where all $t$-dependent corrections and migration effects 
  are applied to the theoretical prediction, which is then fit to the measured raw data.
  
\item The data are split into several sub-samples ordered by time to investigate a potential 
time dependence of the results. 
\end{itemize}   
None of the checks have shown a significant deviation from the nominal results.

\begin{table}[!h]
  \begin{center}
    \begin{tabular}{lcccc}
      \hline \hline
           & Subtraction & Local angle & Lattice & Local subtraction \\ \hline 
$\sigmatot$[mb] & $95.31\pm 0.12$ & $94.96\pm 0.12$ & $95.06\pm 0.12$& $95.03\pm 0.11$\\ 
$B[\GeV^{-2}]$ & $19.62\pm 0.05$ & $19.44\pm 0.05$ & $19.50\pm 0.05$& $19.49\pm 0.05$\\ 
$\chi^2/N_{\mathrm{dof}}$ & 2.8 & 1.07 & 1.2 & 1.4 \\  \hline \hline
    \end{tabular}
  \caption{Fitted values of the total cross section and nuclear slope when only statistical errors are included.}
  \label{tab:fit_stat}
  \end{center}
\end{table}
\subsection{Total inelastic and elastic cross sections}
\label{sec:inelastic}
The total elastic cross section is derived from the nuclear scattering term under the assumption that the
slope $B$ remains constant over the full $t$-range. The Coulomb and interference terms are not taken into account.
With this approximation the differential elastic cross section is reduced to the exponential form:
\begin{eqnarray}
\frac{\mathrm{d}\sigma_{\mathrm{el}}}{\mathrm{d}t} = {\frac{\mathrm{d}\sigma_{\mathrm{el}}}{\mathrm{d}t}}\bigg|_{t=0} \exp(-B|t|) \hspace*{10mm} 
\mbox{with} \hspace*{10mm} {\frac{\mathrm{d}\sigma_{\mathrm{el}}}{\mathrm{d}t}}\bigg|_{t=0} = \sigmatot^2 \frac{1+\rho^2}{16\pi(\hbar c)^2} \: .
\end{eqnarray}  
The differential cross section at the optical point, $|t|\rightarrow 0$, derived from the total cross-section 
fit, is: $\mathrm{d}\sigma_{\mathrm{el}}/\mathrm{d}t|_{t\rightarrow 0} = 474 \pm 4 \; (\mbox{stat.}) \pm 13 \;(\mbox{syst.})$ mb/\GeV$^2$, 
where the systematic uncertainty includes all experimental and extrapolation uncertainties.   
Integrating the parameterized form of the differential cross section over the full $t$-range 
yields the total elastic cross section:
\begin{displaymath}
\sigma_{\mathrm{el}} = 24.00 \pm 0.19 \; (\mbox{stat.}) \pm 0.57 \; (\mbox{syst.}) \; \mbox{mb},  
\end{displaymath}
where the correlation between $\sigmatot$ and $B$, determined from the fit to be 
approximately 40$\%$, is taken into account in the error calculation. 
The measured integrated elastic cross section in the fiducial range from $-t=0.0025$ \GeV$^2$ to $-t=0.38$ \GeV$^2$
corresponds to $90\%$ of the total elastic cross section:
\begin{displaymath}
\sigma_{\mathrm{el}}^{\mathrm{observed}} = 21.66 \pm 0.02 \;(\mbox{stat.}) \pm 0.58 \; (\mbox{syst.}) \; \mbox{mb}.
\end{displaymath} 
The total elastic cross section is used to determine the total inelastic cross section 
by subtraction from the total cross section. The resulting value is:
\begin{displaymath}
\sigma_{\mathrm{inel}} = 71.34 \pm 0.36 \; (\mbox{stat.}) \pm 0.83 \; (\mbox{syst.}) \; \mbox{mb}.
\end{displaymath}
A summary of the derived quantities with uncertainties is given in Table~\ref{tab:inel_syst_summary}.     
\begin{table}
  \begin{center}
    \begin{tabular}{lcccc}
      \hline \hline
           & Optical point [mb/\GeV$^2$] & $\sigma_\mathrm{el}$ [mb]  & $\sigma_\mathrm{inel}$ [mb] & $\sigma_\mathrm{el}^\mathrm{observed}$ [mb] \\ \hline 
Result & 474 & 24.00 & 71.34 & 21.66 \\ 
Statistical error & 4 & 0.19 & 0.36 & 0.02 \\ 
Experimental error& 12 & 0.57 & 0.72 & 0.58 \\
Extrapolation  error& 4 & 0.03 & 0.40 & - \\ \hline
Total error & 13  & 0.60 & 0.90 & 0.58 \\ \hline  \hline
    \end{tabular}
  \caption{Measured values of the optical point, extrapolated elastic cross section, inelastic cross section, and
observed elastic cross section within fiducial cuts.}
  \label{tab:inel_syst_summary}
  \end{center}
\end{table}

\subsection{Model dependence of the nuclear amplitude}
\label{sec:model_dependence}
Traditionally, the nuclear amplitude at small $t$ is parameterized by a single exponential function.
This is the canonical form of the $t$-evolution of the nuclear amplitude  
and was used by several previous experiments \cite{CDF,E811,TOTEM_first} 
to extract the total cross section. 
At larger $t$, approaching the dip around $-t=0.5 \GeV^2$ \cite{TOTEM_larget}, deviations from the 
exponential form are expected. 
In order to assess the impact of using a simple exponential for the nuclear amplitude, 
several alternative forms are investigated.
One of the simplest extensions providing a $t$-dependent slope was used by previous experiments \cite{E710} and discussed in 
Refs.~\cite{WestAndYennie, Block_and_Cahn_curvature}, and consists of an additional term $Ct^2$
in the exponential. Another parameterization was recently proposed \cite{Selyugin}, which considers 
hadron spin non-flip amplitudes contributing to a non-exponential form through an additional term scaling with $\sqrt{-t}$ in 
the exponential. The stochastic vacuum model (SVM) \cite{KFK}, which provides an expression for the Coulomb phase,  
also proposes a parametric form for the differential elastic cross section, assuming constant but possibly different 
slopes for the real and imaginary amplitudes. 
  While the three forms above are to be used in conjunction with the Coulomb amplitude, other parameterizations exist 
with even more model-independent parametric forms, partially absorbing Coulomb effects in the parameterization 
with the use of 
more free parameters. One of these forms (BP) was proposed in Ref.~\cite{PhillipsAndBarger} 
and more recently modified \cite{Fagundes} for the analysis of LHC data, where in total five free parameters are fit. 
Another model-independent parameterization (BSW) with six free parameters, which allow for normalization, offset and slopes separately 
for real and imaginary amplitudes, was suggested  
in Ref.~\cite{BourrelyAndSoffer}.  

Since the deviations from the exponential form are expected 
to increase at larger $t$, the upper limit of the fit range is increased to $-t=0.3$ \GeV$^2$ when fitting the alternative forms. 
Only parametric models have been chosen in order to apply the 
same fit techniques as for the standard analysis.

All alternative forms have at least one more free parameter which improves the quality of the fits
at larger $t$, where best sensitivity for additional parameters is obtained.  
The various fit results using alternative parameterizations are summarized in 
Table~\ref{tab:extrapolation_fit}.
\begin{table}[h!]
  \begin{center}
    \begin{tabular}{lcl}
      \hline \hline
           &  $\sigmatot$[mb] & Reference \\ \hline 
Nominal & $95.35\pm1.30$ & Eq.~\eqref{eq:tgen}\\ 
$Ct^2$ & $95.49\pm1.27$ & Refs.~\cite{WestAndYennie, Block_and_Cahn_curvature}   \\ 
$c\sqrt{-t}$ & $96.03\pm1.31$ & Ref.~\cite{Selyugin}  \\ 
SVM & $94.90\pm1.23$  & Ref.~\cite{KFK}  \\ 
BP & $95.49\pm1.54$ & Refs.~\cite{Fagundes, PhillipsAndBarger}  \\ 
BSW & $95.53\pm1.38$ & Ref.~\cite{BourrelyAndSoffer} \\ \hline \hline
    \end{tabular}
  \caption{Fit results for the total cross section for  
  different parameterizations of the $t$-dependence described in the text. The errors comprise all experimental 
  uncertainties but no theoretical or extrapolation uncertainties.}
  \label{tab:extrapolation_fit}
  \end{center}
\end{table}
The profile fit Eq.~\eqref{eq:profile_fit} including all systematic 
errors was used to evaluate the total cross sections. 
The RMS of the values is in good agreement with the value of 0.4 mb assigned  
to the extrapolation uncertainty obtained with the simple exponential form as described in Section~\ref{sec:sigma_tot_syst}. 
Similar results are obtained for all $t$-reconstruction methods. 

\section{Discussion}
\label{sec:discussion}
The result for the total hadronic cross section presented here, $\sigmatot=95.35 \pm 1.36$ mb, can be compared to the 
most precise value measured by TOTEM, 
in the same LHC fill using a luminosity-dependent analysis, 
$\sigmatot=98.6 \pm 2.2$~mb \cite{TOTEM_second}. Assuming the uncertainties are uncorrelated, the difference 
between the ATLAS and TOTEM values corresponds to 1.3$\sigma$. The uncertainty on the TOTEM result is dominated by the 
luminosity uncertainty of $\pm 4 \%$, giving a $\pm$2 mb contribution to $\sigmatot$ through the square root 
dependence of $\sigmatot$ on luminosity. The measurement reported here profits from a smaller luminosity uncertainty 
of only $\pm2.3\%$.
In subsequent publications \cite{TOTEM_lumindep,TOTEM_inel} TOTEM has used the same data to perform a luminosity-independent 
measurement of the total cross section using a simultaneous determination of elastic and inelastic event yields. 
In addition, TOTEM made a $\rho$-independent measurement without using the optical theorem by summing directly 
the elastic and inelastic cross sections~\cite{TOTEM_lumindep}. The three TOTEM results are consistent with one another.

\begin{figure}[h!]
  \centering
  \includegraphics[width=\textwidth]{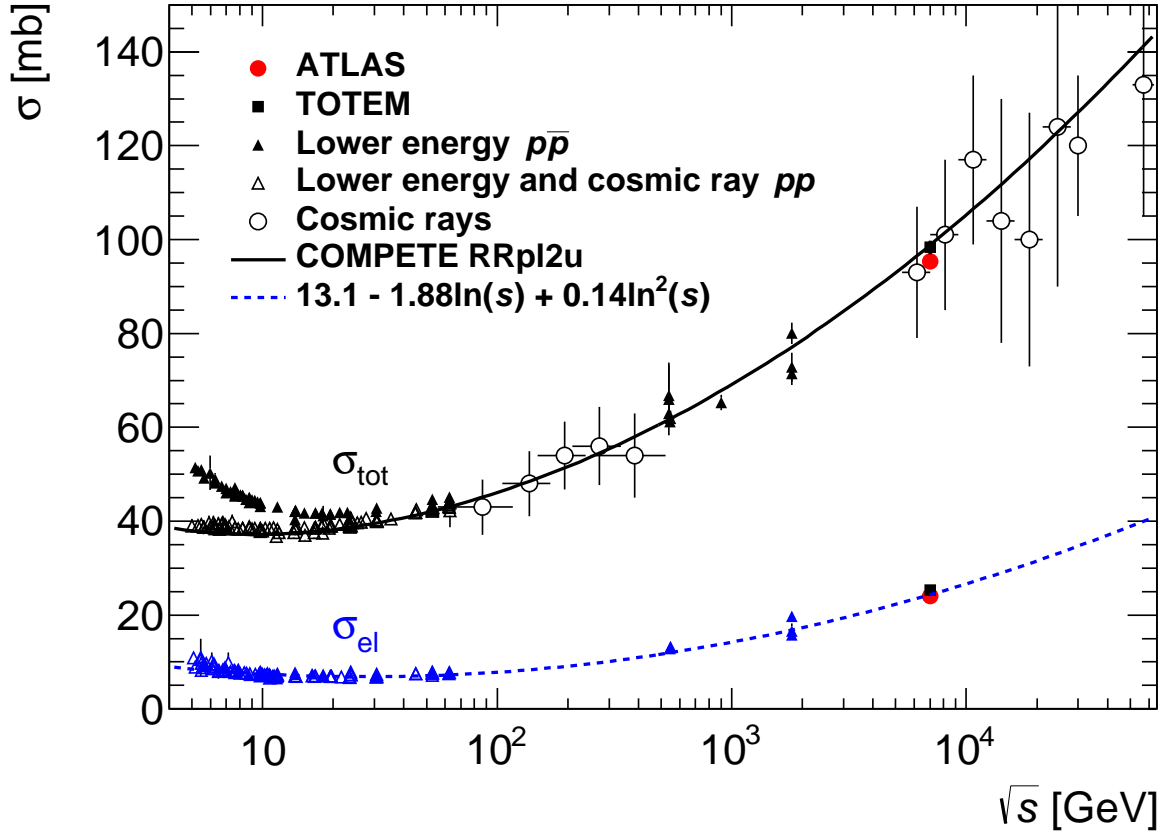}
  \caption{Comparison of total and elastic cross-section measurements presented here with other
 published measurements~\cite{PDG_2012,TOTEM_second,Auger,ARGO-YBJ,AKENO,FlysEye} and model predictions as function of the centre-of-mass energy.} 
  \label{fig:CrossSectionS}
\end{figure}

The results presented here are compared in Fig.~\ref{fig:CrossSectionS} to the  result of TOTEM  and are also compared with 
results of experiments at lower energy~\cite{PDG_2012} and with cosmic ray experiments~\cite{Auger,ARGO-YBJ,AKENO,FlysEye}. 
The measured  total cross section is furthermore compared to the best fit to the energy evolution 
of the total cross section from the COMPETE Collaboration~\cite{compete} assuming  an energy dependence 
of $\ln^2s$. The elastic measurement is in turn compared to a second order polynomial 
fit in $\ln s$ of the elastic cross sections.  The value of $\sigmatot$ reported here is two standard deviations below the COMPETE parameterization.  
Some other models prefer a somewhat 
slower increase  of the total cross section with energy, predicting values below 95 mb, and thus agree slightly 
better with the result reported here~\cite{BlockHalzen,KMR1,Soffer}.

\par
\begin{figure}[h!]
  \centering
  \includegraphics[width=0.8\textwidth]{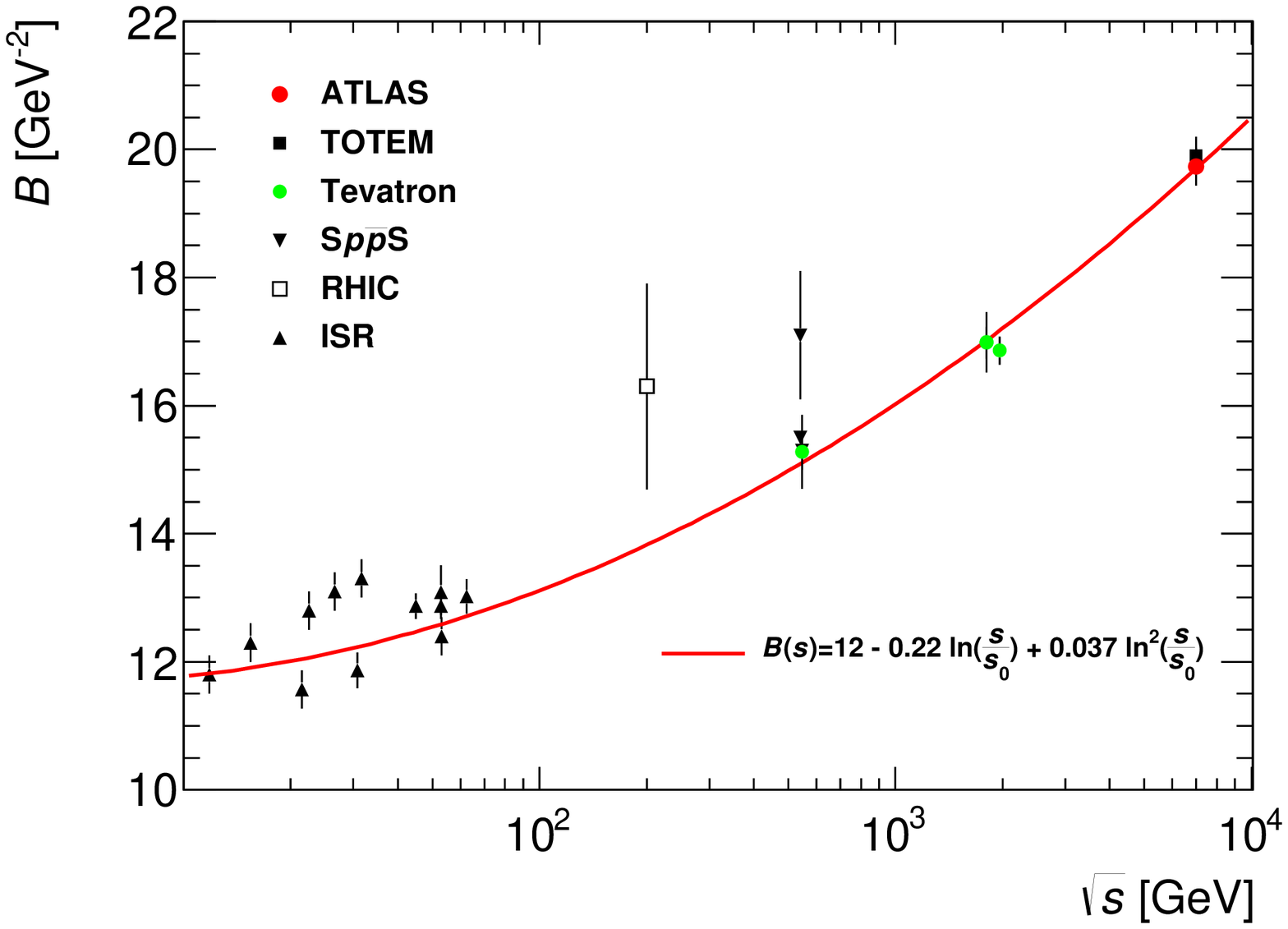}
  \caption{Comparison of the measurement of the nuclear slope $B$ presented here with other
 published measurements at the ISR~\cite{ISR_CR_B,ISR_ACHGT_B,ISR_R_211_B,ISR_R_210}, at the S$p\bar{p}$S~\cite{UA4_B,UA1_B,UA4/2_B}, at RHIC~\cite{PP2PP}, 
at the Tevatron~\cite{CDF,E710_B,D0} and with the measurement from TOTEM~\cite{TOTEM_second} at the LHC. 
The red line shows the calculation from Ref.~\cite{Scheg}, which contains a linear and quadratic term in $\ln s$.} 
\label{fig:BslopeS}
\end{figure}

The value of the nuclear slope parameter $B=19.73 \pm 0.29$ \GeV$^{-2}$ reported here is 
in good agreement with the TOTEM measurement of  $ 19.89 \pm 0.27$ \GeV$^{-2}$ \cite{TOTEM_second}. 
These large values of the $B$-parameter confirm that elastically scattered protons continue
to be confined to a gradually narrowing cone as the energy increases as can be seen from 
Fig.~\ref{fig:BslopeS}. The $B$-parameter is related to the proton radius and as for the total cross section 
an increase can be expected at higher energies. As outlined in Ref.~\cite{Scheg}, the evolution of $B$ from ISR to LHC 
energies is more compatible with a quadratic than a linear dependence in $\ln s$. 

\par
The elastic cross section is measured to be $24.0\pm 0.6$ mb. This is in agreement  with the TOTEM result of $25.4\pm 1.1$ mb within 1.1$\sigma$.
The ratio of the elastic cross section to the total cross section is often taken as a measure of the opacity of the proton.
Measurements shed light on whether the black disc limit of a ratio of 0.5 is being approached. 
It is interesting to note that although there are some small differences between ATLAS and TOTEM 
for the total and elastic cross sections, the ratio $\sigma_{\mathrm{el}}/\sigma_{\mathrm{tot}}$   is very similar. The TOTEM value is 
$\sigma_{\mathrm{el}}/\sigmatot =0.257 \pm0.005$ \cite{TOTEM_lumindep,TOTEM_inel}, while the measurement reported here gives
$\sigma_{\mathrm{el}}/\sigmatot=0.252\pm0.004$. All derived measurements depend on $\sigmatot$ and $B$ and are therefore highly correlated.

\begin{figure}[h!]
  \centering
 \includegraphics[width=0.8\textwidth]{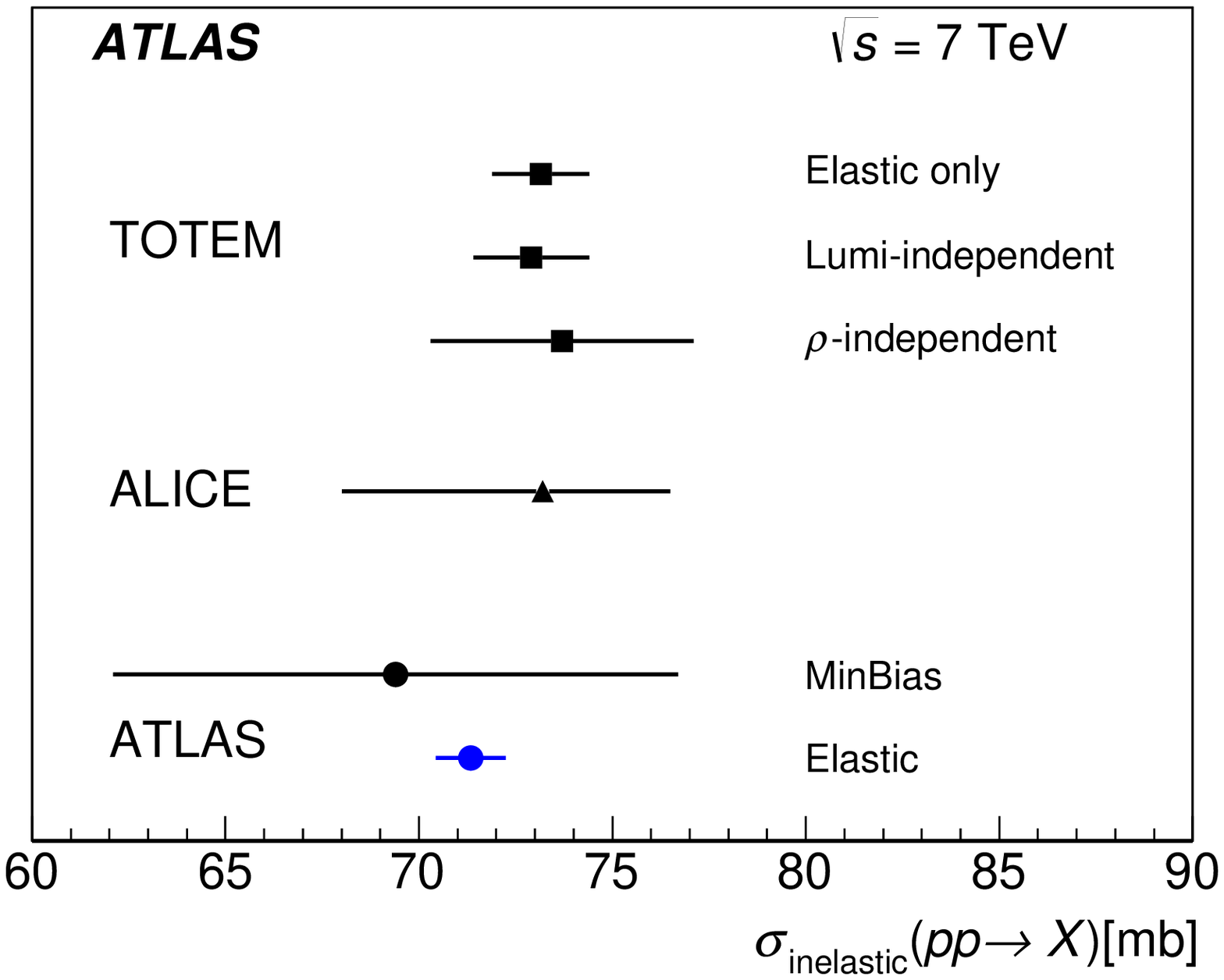}
  \caption{Comparison of the inelastic cross section presented here with the earlier ATLAS measurement~\cite{ATLAS_inel} 
and published data from ALICE~\cite{ALICE_inel} and TOTEM~\cite{TOTEM_second,TOTEM_lumindep,TOTEM_inel}.} 
  \label{fig:inelastic}
\end{figure}

Finally, various total inelastic cross-section measurements, either using only elastic data as in the present 
analysis and in the corresponding analysis of TOTEM \cite{TOTEM_second}, or using elastic and 
inelastic data in the luminosity-independent method \cite{TOTEM_lumindep}, are compared in 
Fig.~\ref{fig:inelastic} to direct measurements by TOTEM~\cite{TOTEM_inel},  
ALICE~\cite{ALICE_inel} and an  ATLAS~\cite{ATLAS_inel} result based upon an analysis of minimum bias events. In general, the direct measurements have 
a larger uncertainty because of the model dependence when extrapolating to the
unobservable cross section 
at low diffractive masses. A measurement from CMS~\cite{CMS_inel} was not extrapolated to the total inelastic cross 
section and is therefore not included in Fig.~\ref{fig:inelastic}. 
The measurement of $\sigma_{\mathrm{inel}}=71.34 \pm 0.90$~mb  reported here improves considerably the previous 
measurement of ATLAS: $\sigma_{\mathrm{inel}}=69.4 \pm 7.3$~mb ~\cite{ATLAS_inel}.     

As mentioned above, the previous ATLAS result for the inelastic cross section has a large uncertainty because of 
the uncertainty in the extrapolation to low diffractive masses outside the fiducial region. 
The fiducial region covers only dissociative processes with the diffractive invariant mass, $M_X$, larger than 15.7~\GeV.
However, the precision of the measurement in the fiducial region is good, with an uncertainty on the fiducial cross section 
roughly three times smaller than on the extrapolated inelastic cross section. Thus by taking the difference of the present measurement 
of the total inelastic cross section and the previous measurement by ATLAS in the fiducial region 
the size of the cross section for invariant masses below 15.7 \GeV\ can be estimated and compared with existing models. 
The cross section measured by ATLAS corresponding to masses above 15.7 \GeV\ is $60.3 \pm 2.1$ mb~\cite {ATLAS_inel}. 
Thus the cross section in which the dissociative systems have an invariant mass $M_X$ less than 15.7 \GeV\ 
is estimated to be $ 11.0 \pm 2.3$ mb. Both PYTHIA and PHOJET~\cite {PHOJET1, PHOJET2} predict significantly 
lower contributions to the inelastic cross section in this range of invariant masses, approximately 6 mb and 3 mb  respectively.  
The model of Ref.~\cite{KMR1} gives a better description of the data with values in the range 11--14 mb (see discussion in Ref.~\cite{Rapgap}).

\section{Summary}
\label{sec:conclusion}
In this paper a measurement of the elastic $pp$ cross section and the determination of the total cross section  using the optical theorem at $\sqrt{s}=7 \TeV$ by the ATLAS experiment 
at the LHC with the ALFA sub-detector is presented. The data were recorded in 2011 during a special 
run with high-$\betastar$ optics, where an integrated luminosity of 
$80$ $\mu$b$^{-1}$ was accumulated. The analysis uses data-driven methods to determine relevant 
beam optics parameters, event reconstruction efficiency and to tune the simulation. A key element of 
this analysis is the determination of the effective beam optics, which takes into account measurements 
from ALFA that are sensitive to ratios of transport matrix elements and calibration uncertainties of the quadrupoles. 
A detailed evaluation of the associated systematic uncertainties includes  the comparison of 
different $t$-reconstruction methods that are sensitive to different transport matrix elements.  
A dedicated effort was made to determine the absolute luminosity for this run while
taking into account the  special conditions with a very low number of interactions per bunch crossing. 
From a fit to the differential elastic cross section, using the optical theorem, the total cross section is determined to be:
\begin{eqnarray*}
\sigmatot(pp\rightarrow X) & = &  \mbox{95.35} \; \pm \mbox{0.38} \; (\mbox{stat.}) \pm \mbox{1.25} \; (\mbox{exp.})  \pm \mbox{0.37} \; (\mbox{extr.})  \; \mbox{mb} \; ,
\end{eqnarray*}
where the first error is statistical, the second accounts for all experimental systematic uncertainties and the last 
is related to uncertainties on the extrapolation to $|t|\rightarrow 0$. 
The experimental systematic uncertainty is dominated by the uncertainty on the luminosity and on the 
nominal beam energy. 
Further experimental uncertainties including those associated with beam optics and detector modelling 
contribute less. Alternative models for the nuclear amplitude parameterization with non-exponential contributions 
were investigated using an extended fit range and the resulting values of the total cross section range from 94.9 mb 
to 96.0 mb.        

In addition the slope of the elastic differential cross section at small $t$ was determined to be:
\begin{equation*}
B = \mbox{19.73} \; \pm \mbox{0.14} \; (\mbox{stat.})
\pm \mbox{0.26} \; (\mbox{syst.}) \;  \mbox{GeV}^{-2} \; . 
\end{equation*}
From the fitted parameterization of the elastic cross section the total elastic cross section is extracted:
\begin{equation*}
\sigma_{\mathrm{el}}(pp\rightarrow pp) =  \mbox{24.00} \; \pm \mbox{0.19} \; (\mbox{stat.})
\pm \mbox{0.57} \; (\mbox{syst.}) \; \mbox{mb} \; , 
\end{equation*}
and by subtraction from the total cross section the inelastic cross section is determined to be:
\begin{equation*}
\sigma_{\mathrm{inel}} =  \mbox{71.34} \; \pm \mbox{0.36} \; (\mbox{stat.})
\pm \mbox{0.83} \; (\mbox{syst.})\;  \mbox{mb} \; , 
\end{equation*}
which is significantly more precise than the previous direct ATLAS measurement.


\section{Acknowledgements}

We thank CERN for the very successful operation of the LHC, as well as the
support staff from our institutions without whom ATLAS could not be
operated efficiently.

We acknowledge the support of ANPCyT, Argentina; YerPhI, Armenia; ARC,
Australia; BMWFW and FWF, Austria; ANAS, Azerbaijan; SSTC, Belarus; CNPq and FAPESP,
Brazil; NSERC, NRC and CFI, Canada; CERN; CONICYT, Chile; CAS, MOST and NSFC,
China; COLCIENCIAS, Colombia; MSMT CR, MPO CR and VSC CR, Czech Republic;
DNRF, DNSRC and Lundbeck Foundation, Denmark; EPLANET, ERC and NSRF, European Union;
IN2P3-CNRS, CEA-DSM/IRFU, France; GNSF, Georgia; BMBF, DFG, HGF, MPG and AvH
Foundation, Germany; GSRT and NSRF, Greece; ISF, MINERVA, GIF, I-CORE and Benoziyo Center,
Israel; INFN, Italy; MEXT and JSPS, Japan; CNRST, Morocco; FOM and NWO,
Netherlands; BRF and RCN, Norway; MNiSW and NCN, Poland; GRICES and FCT, Portugal; MNE/IFA, Romania; MES of Russia and ROSATOM, Russian Federation; JINR; MSTD,
Serbia; MSSR, Slovakia; ARRS and MIZ\v{S}, Slovenia; DST/NRF, South Africa;
MINECO, Spain; SRC and Wallenberg Foundation, Sweden; SER, SNSF and Cantons of
Bern and Geneva, Switzerland; NSC, Taiwan; TAEK, Turkey; STFC, the Royal
Society and Leverhulme Trust, United Kingdom; DOE and NSF, United States of
America.

The crucial computing support from all WLCG partners is acknowledged
gratefully, in particular from CERN and the ATLAS Tier-1 facilities at
TRIUMF (Canada), NDGF (Denmark, Norway, Sweden), CC-IN2P3 (France),
KIT/GridKA (Germany), INFN-CNAF (Italy), NL-T1 (Netherlands), PIC (Spain),
ASGC (Taiwan), RAL (UK) and BNL (USA) and in the Tier-2 facilities
worldwide.

\bibliographystyle{model1a-num-names}
\bibliography{elastics7}

\begin{thebibliography}{77}
\expandafter\ifx\csname natexlab\endcsname\relax\def\natexlab#1{#1}\fi
\providecommand{\url}[1]{\texttt{#1}}
\providecommand{\href}[2]{#2}
\providecommand{\path}[1]{#1}
\providecommand{\DOIprefix}{doi:}
\providecommand{\ArXivprefix}{arXiv:}
\providecommand{\URLprefix}{URL: }
\providecommand{\Pubmedprefix}{pmid:}
\providecommand{\doi}[1]{\href{http://dx.doi.org/#1}{\path{#1}}}
\providecommand{\Pubmed}[1]{\href{pmid:#1}{\path{#1}}}
\providecommand{\bibinfo}[2]{#2}
\ifx\xfnm\relax \def\xfnm[#1]{\unskip,\space#1}\fi
\bibitem[{Froissart(1961)}]{froissart}
\bibinfo{author}{M.~Froissart}, \bibinfo{journal}{Phys. Rev.}
  \bibinfo{volume}{123} (\bibinfo{year}{1961}) \bibinfo{pages}{1053}.
\bibitem[{Martin(1966)}]{martin1}
\bibinfo{author}{A.~Martin}, \bibinfo{journal}{Il Nuovo Cimento A Series 10}
  \bibinfo{volume}{42} (\bibinfo{year}{1966}) \bibinfo{pages}{930}.
\bibitem[{Amendolia et~al.(1973)}]{Pis1}
\bibinfo{author}{S.~R. Amendolia}, et~al., \bibinfo{journal}{Phys. Lett.}
  \bibinfo{volume}{B 44} (\bibinfo{year}{1973}) \bibinfo{pages}{119}.
\bibitem[{Amaldi et~al.(1973)}]{Amaldi2}
\bibinfo{author}{U.~Amaldi}, et~al., \bibinfo{journal}{Phys. Lett.}
  \bibinfo{volume}{B 44} (\bibinfo{year}{1973}) \bibinfo{pages}{112}.
\bibitem[{Bozzo et~al.(1984)}]{UA4}
\bibinfo{author}{M.~Bozzo}, et~al. (\bibinfo{collaboration}{UA4
  Collaboration}), \bibinfo{journal}{Phys. Lett.} \bibinfo{volume}{B 147}
  (\bibinfo{year}{1984}) \bibinfo{pages}{392}.
\bibitem[{Augier et~al.(1995)}]{UA42}
\bibinfo{author}{C.~Augier}, et~al. (\bibinfo{collaboration}{UA4/2
  Collaboration}), \bibinfo{journal}{Phys. Lett.} \bibinfo{volume}{B 344}
  (\bibinfo{year}{1995}) \bibinfo{pages}{451}.
\bibitem[{Amos et~al.(1990)}]{E710A}
\bibinfo{author}{N.~A. Amos}, et~al. (\bibinfo{collaboration}{E710
  Collaboration}), \bibinfo{journal}{Phys. Lett.} \bibinfo{volume}{B 243}
  (\bibinfo{year}{1990}) \bibinfo{pages}{158}.
\bibitem[{Avila et~al.(1999)}]{E811}
\bibinfo{author}{C.~Avila}, et~al. (\bibinfo{collaboration}{E811
  Collaboration}), \bibinfo{journal}{Phys. Lett.} \bibinfo{volume}{B 445}
  (\bibinfo{year}{1999}) \bibinfo{pages}{419}.
\bibitem[{Abe et~al.(1994)}]{CDF}
\bibinfo{author}{F.~Abe}, et~al. (\bibinfo{collaboration}{CDF Collaboration}),
  \bibinfo{journal}{Phys. Rev.} \bibinfo{volume}{D 50} (\bibinfo{year}{1994})
  \bibinfo{pages}{5518}.
\bibitem[{Antchev et~al.(2011)}]{TOTEM_first}
\bibinfo{author}{G.~Antchev}, et~al. (\bibinfo{collaboration}{TOTEM
  Collaboration}), \bibinfo{journal}{Europhys. Lett.} \bibinfo{volume}{96}
  (\bibinfo{year}{2011}) \bibinfo{pages}{21002}.
  \href{http://arxiv.org/abs/1110.1395}{\tt arXiv:1110.1395}.
\bibitem[{Antchev et~al.(2013)}]{TOTEM_second}
\bibinfo{author}{G.~Antchev}, et~al. (\bibinfo{collaboration}{TOTEM
  Collaboration}), \bibinfo{journal}{Europhys. Lett.} \bibinfo{volume}{101}
  (\bibinfo{year}{2013}) \bibinfo{pages}{21002}.
\bibitem[{{ATLAS Collaboration}(2008)}]{atlas1}
\bibinfo{author}{{ATLAS Collaboration}}, \bibinfo{journal}{JINST}
  \bibinfo{volume}{3} (\bibinfo{year}{2008}) \bibinfo{pages}{S08003}.
\bibitem[{van~der Meer(1968)}]{svdm}
\bibinfo{author}{S.~van~der Meer}, \bibinfo{journal}{ISR-PO-68-31}
  (\bibinfo{year}{1968}) \bibinfo{pages}{http://cds.cern.ch/record/296752}.
\bibitem[{{ATLAS Collaboration}(2011)}]{ATLAS_inel}
\bibinfo{author}{{ATLAS Collaboration}}, \bibinfo{journal}{Nature Commun.}
  \bibinfo{volume}{2} (\bibinfo{year}{2011}) \bibinfo{pages}{463}.
  \href{http://arxiv.org/abs/1104.0326}{\tt arXiv:1104.0326}.
\bibitem[{Antchev et~al.(2013{\natexlab{a}})}]{totem1}
\bibinfo{author}{G.~Antchev}, et~al. (\bibinfo{collaboration}{TOTEM
  Collaboration}), \bibinfo{journal}{Phys. Rev. Lett.} \bibinfo{volume}{111}
  (\bibinfo{year}{2013}{\natexlab{a}}) \bibinfo{pages}{012001}.
  \href{http://arxiv.org/abs/1308.6722}{\tt arXiv:1308.6722}.
\bibitem[{Antchev et~al.(2013{\natexlab{b}})}]{TOTEM_lumindep}
\bibinfo{author}{G.~Antchev}, et~al. (\bibinfo{collaboration}{TOTEM
  Collaboration}), \bibinfo{journal}{Europhys. Lett.} \bibinfo{volume}{101}
  (\bibinfo{year}{2013}{\natexlab{b}}) \bibinfo{pages}{21004}.
\bibitem[{Anghinolfi et~al.(2007)}]{testbeam1}
\bibinfo{author}{F.~Anghinolfi}, et~al., \bibinfo{journal}{JINST}
  \bibinfo{volume}{2} (\bibinfo{year}{2007}) \bibinfo{pages}{P07004}.
  \href{http://arxiv.org/abs/0706.3316}{\tt arXiv:0706.3316}.
\bibitem[{Ask et~al.(2006)}]{testbeam2}
\bibinfo{author}{S.~Ask}, et~al., \bibinfo{journal}{Nucl. Instrum. Meth.}
  \bibinfo{volume}{A 568} (\bibinfo{year}{2006}) \bibinfo{pages}{588}.
  \href{http://arxiv.org/abs/physics/0605127}{\tt arXiv:physics/0605127}.
\bibitem[{Barrillon et~al.(2010)}]{MAROC1}
\bibinfo{author}{P.~Barrillon}, et~al., \bibinfo{journal}{Nucl. Instrum. Meth.}
  \bibinfo{volume}{A 623} (\bibinfo{year}{2010}) \bibinfo{pages}{463}.
\bibitem[{Blin et~al.(2010)Blin, Barrillon, and de~La~Taille}]{MAROC2}
\bibinfo{author}{S.~Blin}, \bibinfo{author}{P.~Barrillon},
  \bibinfo{author}{C.~de~La~Taille}, \bibinfo{journal}{IEEE Nucl. Sci. Symp
  .Conf. Rec.} \bibinfo{volume}{2010} (\bibinfo{year}{2010})
  \bibinfo{pages}{1690--1693}.
\bibitem[{Burkhardt et~al.(2012)}]{Note90m}
\bibinfo{author}{H.~Burkhardt}, et~al., \bibinfo{journal}{Conf. Proc.}
  \bibinfo{volume}{C1205201} (\bibinfo{year}{2012}) \bibinfo{pages}{130}.
\bibitem[{Burkhardt et~al.(2011)Burkhardt, Cavalier, Puzo, and
  Peskov}]{HBLumiday11}
\bibinfo{author}{H.~Burkhardt}, \bibinfo{author}{S.~Cavalier},
  \bibinfo{author}{P.~Puzo}, \bibinfo{author}{A.~Peskov},
  \bibinfo{journal}{Conf. Proc.} \bibinfo{volume}{C110904}
  (\bibinfo{year}{2011}) \bibinfo{pages}{1798}.
\bibitem[{Bethe(1958)}]{Bethe}
\bibinfo{author}{H.~A. Bethe}, \bibinfo{journal}{Ann. Phys.}
  \bibinfo{volume}{3} (\bibinfo{year}{1958}) \bibinfo{pages}{190}.
\bibitem[{West and Yennie(1968)}]{WestAndYennie}
\bibinfo{author}{G.~B. West}, \bibinfo{author}{D.~R. Yennie},
  \bibinfo{journal}{Phys. Rev.} \bibinfo{volume}{172} (\bibinfo{year}{1968})
  \bibinfo{pages}{1413}.
\bibitem[{Cahn(1982)}]{Cahn}
\bibinfo{author}{R.~N. Cahn}, \bibinfo{journal}{Z. Phys.} \bibinfo{volume}{C
  15} (\bibinfo{year}{1982}) \bibinfo{pages}{253}.
\bibitem[{Cudell et~al.(2002)}]{compete}
\bibinfo{author}{J.~Cudell}, et~al. (\bibinfo{collaboration}{COMPETE
  Collaboration}), \bibinfo{journal}{Phys. Rev. Lett.} \bibinfo{volume}{89}
  (\bibinfo{year}{2002}) \bibinfo{pages}{201801}.
  \href{http://arxiv.org/abs/hep-ph/0206172}{\tt arXiv:hep-ph/0206172}.
\bibitem[{Yao et~al.(2006)}]{PDG_2005}
\bibinfo{author}{W.~M. Yao}, et~al. (\bibinfo{collaboration}{Particle Data
  Group}), \bibinfo{journal}{J. Phys.} \bibinfo{volume}{G 33}
  (\bibinfo{year}{2006}) \bibinfo{pages}{1}.
\bibitem[{Alkin et~al.(2012)Alkin, Cudell, and Martinov}]{cudell_dispersion}
\bibinfo{author}{A.~Alkin}, \bibinfo{author}{J.~R. Cudell},
  \bibinfo{author}{E.~Martinov}, \bibinfo{journal}{Few Body Syst.}
  \bibinfo{volume}{53} (\bibinfo{year}{2012}) \bibinfo{pages}{87}.
  \href{http://arxiv.org/abs/1109.1306}{\tt arXiv:1109.1306}.
\bibitem[{Beringer et~al.(2012)}]{PDG_2012}
\bibinfo{author}{J.~Beringer}, et~al. (\bibinfo{collaboration}{Particle Data
  Group}), \bibinfo{journal}{Phys. Rev.} \bibinfo{volume}{D 86}
  (\bibinfo{year}{2012}) \bibinfo{pages}{010001}.
\bibitem[{Menon and Silva(2013)}]{menon_silva}
\bibinfo{author}{M.~J. Menon}, \bibinfo{author}{P.~V. R.~G. Silva},
  \bibinfo{journal}{J. Phys.} \bibinfo{volume}{G 40} (\bibinfo{year}{2013})
  \bibinfo{pages}{125001}. \href{http://arxiv.org/abs/1305.2947}{\tt
  arXiv:1305.2947}, \bibinfo{note}{erratum-ibid. G41 (2014) 019501}.
\bibitem[{Kohara et~al.(2013)Kohara, Ferreira, and Kodama}]{KFK}
\bibinfo{author}{A.~K. Kohara}, \bibinfo{author}{E.~Ferreira},
  \bibinfo{author}{T.~Kodama}, \bibinfo{journal}{Eur. Phys. J.}
  \bibinfo{volume}{C 73} (\bibinfo{year}{2013}) \bibinfo{pages}{2326}.
  \href{http://arxiv.org/abs/1212.3652}{\tt arXiv:1212.3652}.
\bibitem[{Simon et~al.(1980)Simon, Schmitt, Borkowski, and Walther}]{Simon_FF}
\bibinfo{author}{G.~G. Simon}, \bibinfo{author}{C.~Schmitt},
  \bibinfo{author}{F.~Borkowski}, \bibinfo{author}{V.~H. Walther},
  \bibinfo{journal}{Nucl. Phys.} \bibinfo{volume}{A 333} (\bibinfo{year}{1980})
  \bibinfo{pages}{381}.
\bibitem[{Bernauer et~al.(2014)}]{A1}
\bibinfo{author}{J.~C. Bernauer}, et~al. (\bibinfo{collaboration}{A1
  Collaboration}), \bibinfo{journal}{Phys. Rev.} \bibinfo{volume}{C 90}
  (\bibinfo{year}{2014}) \bibinfo{pages}{015206}.
  \href{http://arxiv.org/abs/1307.6227}{\tt arXiv:1307.6227}.
\bibitem[{Sj\"{o}strand et~al.(2008)Sj\"{o}strand, Mrenna, and Skands}]{PYTHIA}
\bibinfo{author}{T.~Sj\"{o}strand}, \bibinfo{author}{S.~Mrenna},
  \bibinfo{author}{P.~Skands}, \bibinfo{journal}{Comput. Phys. Commun.}
  \bibinfo{volume}{178} (\bibinfo{year}{2008}) \bibinfo{pages}{852}.
  \href{http://arxiv.org/abs/0710.3820}{\tt arXiv:0710.3820}.
\bibitem[{Sj\"{o}strand et~al.(2006)Sj\"{o}strand, Mrenna, and
  Skands}]{PYTHIA6}
\bibinfo{author}{T.~Sj\"{o}strand}, \bibinfo{author}{S.~Mrenna},
  \bibinfo{author}{P.~Skands}, \bibinfo{journal}{JHEP} \bibinfo{volume}{05}
  (\bibinfo{year}{2006}) \bibinfo{pages}{026}.
  \href{http://arxiv.org/abs/hep-ph/0603175}{\tt arXiv:hep-ph/0603175}.
\bibitem[{{CERN Accelerator Beam Physics Group}(2014)}]{madx}
\bibinfo{author}{{CERN Accelerator Beam Physics Group}}, \bibinfo{title}{Mad -
  methodical accelerator design},
  \bibinfo{howpublished}{{http://mad.web.cern.ch/mad/}}, \bibinfo{year}{2014}.
\bibitem[{Allison et~al.(2006)}]{GEANT41}
\bibinfo{author}{J.~Allison}, et~al., \bibinfo{journal}{IEEE Trans. Nucl. Sci.}
  \bibinfo{volume}{53} (\bibinfo{year}{2006}) \bibinfo{pages}{270}.
\bibitem[{Agostinelli et~al.(2003)}]{GEANT42}
\bibinfo{author}{S.~Agostinelli}, et~al. (\bibinfo{collaboration}{GEANT4}),
  \bibinfo{journal}{Nucl. Instrum. Meth.} \bibinfo{volume}{A 506}
  (\bibinfo{year}{2003}) \bibinfo{pages}{250}.
\bibitem[{Baud et~al.(2013)}]{Wire_Scan}
\bibinfo{author}{G.~Baud}, et~al., \bibinfo{journal}{CERN-ACC-2013-0308}
  (\bibinfo{year}{2013}) \bibinfo{pages}{http://cds.cern.ch/record/1638354}.
\bibitem[{Malaescu(2011)}]{IDS}
\bibinfo{author}{B.~Malaescu}, \bibinfo{journal}{CERN-PH-EP-2011-111}
  (\bibinfo{year}{2011}) \bibinfo{pages}{http://cds.cern.ch/record/1365693}.
  \href{http://arxiv.org/abs/1106.3107}{\tt arXiv:1106.3107}.
\bibitem[{H\"{o}cker and Kartvelishvili(1996)}]{SVD}
\bibinfo{author}{A.~H\"{o}cker}, \bibinfo{author}{V.~G. Kartvelishvili},
  \bibinfo{journal}{Nucl. Instrum. Meth.} \bibinfo{volume}{A 372}
  (\bibinfo{year}{1996}) \bibinfo{pages}{469}.
  \href{http://arxiv.org/abs/hep-ph/9509307}{\tt arXiv:hep-ph/9509307}.
\bibitem[{{ATLAS Collaboration}(2013)}]{LumiPaper2011}
\bibinfo{author}{{ATLAS Collaboration}}, \bibinfo{journal}{Eur. Phys. J.}
  \bibinfo{volume}{C 73} (\bibinfo{year}{2013}) \bibinfo{pages}{2518}.
  \href{http://arxiv.org/abs/1302.4393}{\tt arXiv:1302.4393}.
\bibitem[{Wenninger(2013)}]{Wenninger}
\bibinfo{author}{J.~Wenninger}, \bibinfo{title}{{Energy Calibration of the LHC
  Beams at 4 TeV}}, \bibinfo{howpublished}{{CERN-ATS-2013-040,
  http://cds.cern.ch/record/1546734}}, \bibinfo{year}{2013}.
\bibitem[{{The Durham HepData Project}(2014)}]{HepData}
\bibinfo{author}{{The Durham HepData Project}},
  \bibinfo{howpublished}{http://hepdata.cedar.ac.uk/}, \bibinfo{year}{2014}.
\bibitem[{Blobel(2003)}]{profile}
\bibinfo{author}{V.~Blobel}, \bibinfo{journal}{{eConf}} \bibinfo{volume}{{C
  030908}} (\bibinfo{year}{2003}) \bibinfo{pages}{{MOET002.
  http://www.slac.stanford.edu/econf/C030908/proceedings.html}}.
\bibitem[{Khoze et~al.(2000)Khoze, Martin, and Ryskin}]{per_theory}
\bibinfo{author}{V.~A. Khoze}, \bibinfo{author}{A.~D. Martin},
  \bibinfo{author}{M.~G. Ryskin}, \bibinfo{journal}{Eur. Phys. J.}
  \bibinfo{volume}{C 18} (\bibinfo{year}{2000}) \bibinfo{pages}{167}.
  \href{http://arxiv.org/abs/hep-ph/0007359}{\tt arXiv:hep-ph/0007359}.
\bibitem[{Antchev et~al.(2011)}]{TOTEM_larget}
\bibinfo{author}{G.~Antchev}, et~al. (\bibinfo{collaboration}{TOTEM
  Collaboration}), \bibinfo{journal}{Europhys. Lett.} \bibinfo{volume}{95}
  (\bibinfo{year}{2011}) \bibinfo{pages}{41001}.
  \href{http://arxiv.org/abs/1110.1385}{\tt arXiv:1110.1385}.
\bibitem[{Amos et~al.(1990)}]{E710}
\bibinfo{author}{N.~A. Amos}, et~al. (\bibinfo{collaboration}{E710
  Collaboration}), \bibinfo{journal}{Phys. Lett.} \bibinfo{volume}{B 247}
  (\bibinfo{year}{1990}) \bibinfo{pages}{127}.
\bibitem[{Block and Cahn(1990)}]{Block_and_Cahn_curvature}
\bibinfo{author}{M.~M. Block}, \bibinfo{author}{R.~N. Cahn},
  \bibinfo{journal}{Czech. J. Phys.} \bibinfo{volume}{40}
  (\bibinfo{year}{1990}) \bibinfo{pages}{164}.
\bibitem[{Selyugin(2014)}]{Selyugin}
\bibinfo{author}{O.~V. Selyugin}, \bibinfo{journal}{Nucl. Phys.}
  \bibinfo{volume}{A 922} (\bibinfo{year}{2014}) \bibinfo{pages}{180}.
  \href{http://arxiv.org/abs/1312.1271}{\tt arXiv:1312.1271}.
\bibitem[{Phillips and Barger(1973)}]{PhillipsAndBarger}
\bibinfo{author}{R.~J.~N. Phillips}, \bibinfo{author}{V.~D. Barger},
  \bibinfo{journal}{Phys. Lett.} \bibinfo{volume}{B 46} (\bibinfo{year}{1973})
  \bibinfo{pages}{412}.
\bibitem[{Fagundes et~al.(2013)}]{Fagundes}
\bibinfo{author}{D.~A. Fagundes}, et~al., \bibinfo{journal}{Phys. Rev.}
  \bibinfo{volume}{D 88} (\bibinfo{year}{2013}) \bibinfo{pages}{094019}.
  \href{http://arxiv.org/abs/1306.0452}{\tt arXiv:1306.0452}.
\bibitem[{Bourrely et~al.(2011)Bourrely, Soffer, and Wu}]{BourrelyAndSoffer}
\bibinfo{author}{C.~Bourrely}, \bibinfo{author}{J.~Soffer},
  \bibinfo{author}{T.~T. Wu}, \bibinfo{journal}{Eur. Phys. J.}
  \bibinfo{volume}{C 71} (\bibinfo{year}{2011}) \bibinfo{pages}{1601}.
  \href{http://arxiv.org/abs/1011.1756}{\tt arXiv:1011.1756}.
\bibitem[{Antchev et~al.(2013)}]{TOTEM_inel}
\bibinfo{author}{G.~Antchev}, et~al. (\bibinfo{collaboration}{TOTEM
  Collaboration}), \bibinfo{journal}{Europhys. Lett.} \bibinfo{volume}{101}
  (\bibinfo{year}{2013}) \bibinfo{pages}{21003}.
\bibitem[{Abreu et~al.(2012)}]{Auger}
\bibinfo{author}{P.~Abreu}, et~al. (\bibinfo{collaboration}{Pierre Auger
  Collaboration}), \bibinfo{journal}{Phys. Rev. Lett.} \bibinfo{volume}{109}
  (\bibinfo{year}{2012}) \bibinfo{pages}{062002}.
  \href{http://arxiv.org/abs/1208.1520}{\tt arXiv:1208.1520}.
\bibitem[{Aielli et~al.(2009)}]{ARGO-YBJ}
\bibinfo{author}{G.~Aielli}, et~al. (\bibinfo{collaboration}{ARGO-YBJ
  Collaboration}), \bibinfo{journal}{Phys. Rev.} \bibinfo{volume}{D 80}
  (\bibinfo{year}{2009}) \bibinfo{pages}{092004}.
  \href{http://arxiv.org/abs/0904.4198}{\tt arXiv:0904.4198}.
\bibitem[{Honda et~al.(1993)}]{AKENO}
\bibinfo{author}{M.~Honda}, et~al., \bibinfo{journal}{Phys. Rev. Lett.}
  \bibinfo{volume}{70} (\bibinfo{year}{1993}) \bibinfo{pages}{525}.
\bibitem[{Baltrusaitis et~al.(1984)}]{FlysEye}
\bibinfo{author}{R.~M. Baltrusaitis}, et~al., \bibinfo{journal}{Phys. Rev.
  Lett.} \bibinfo{volume}{52} (\bibinfo{year}{1984}) \bibinfo{pages}{1380}.
\bibitem[{Block and Halzen(2012)}]{BlockHalzen}
\bibinfo{author}{M.~M. Block}, \bibinfo{author}{F.~Halzen},
  \bibinfo{journal}{Phys. Rev.} \bibinfo{volume}{D 86} (\bibinfo{year}{2012})
  \bibinfo{pages}{051504}. \href{http://arxiv.org/abs/1208.4086}{\tt
  arXiv:1208.4086}.
\bibitem[{Ryskin et~al.(2011)Ryskin, Martin, and Khoze}]{KMR1}
\bibinfo{author}{M.~G. Ryskin}, \bibinfo{author}{A.~D. Martin},
  \bibinfo{author}{V.~A. Khoze}, \bibinfo{journal}{Eur. Phys. J.}
  \bibinfo{volume}{C 71} (\bibinfo{year}{2011}) \bibinfo{pages}{1617}.
  \href{http://arxiv.org/abs/1102.2844}{\tt arXiv:1102.2844}.
\bibitem[{J.Soffer(2013)}]{Soffer}
\bibinfo{author}{J.Soffer}, \bibinfo{journal}{AIP Conf. Proc.}
  \bibinfo{volume}{1523} (\bibinfo{year}{2013}) \bibinfo{pages}{115}.
  \href{http://arxiv.org/abs/1302.5045}{\tt arXiv:1302.5045}.
\bibitem[{Schegelsky and Ryskin(2012)}]{Scheg}
\bibinfo{author}{V.~A. Schegelsky}, \bibinfo{author}{M.~G. Ryskin},
  \bibinfo{journal}{Phys. Rev.} \bibinfo{volume}{D 85} (\bibinfo{year}{2012})
  \bibinfo{pages}{094024}. \href{http://arxiv.org/abs/1112.3243}{\tt
  arXiv:1112.3243}.
\bibitem[{Amaldi et~al.(1976)}]{ISR_CR_B}
\bibinfo{author}{U.~Amaldi}, et~al.
  (\bibinfo{collaboration}{CERN-Pisa-Rome-Stony Brook Collaboration}),
  \bibinfo{journal}{Phys. Lett.} \bibinfo{volume}{B 62} (\bibinfo{year}{1976})
  \bibinfo{pages}{460}.
\bibitem[{Barbiellini et~al.(1972)}]{ISR_ACHGT_B}
\bibinfo{author}{G.~Barbiellini}, et~al., \bibinfo{journal}{Phys. Lett.}
  \bibinfo{volume}{B 39} (\bibinfo{year}{1972}) \bibinfo{pages}{663}.
\bibitem[{Amos et~al.(1985)}]{ISR_R_211_B}
\bibinfo{author}{N.~A. Amos}, et~al., \bibinfo{journal}{Nucl. Phys.}
  \bibinfo{volume}{B 262} (\bibinfo{year}{1985}) \bibinfo{pages}{689}.
\bibitem[{Ambrosio et~al.(1982)}]{ISR_R_210}
\bibinfo{author}{M.~Ambrosio}, et~al.
  (\bibinfo{collaboration}{CERN-Naples-Pisa-Stony Brook Collaboration}),
  \bibinfo{journal}{Phys. Lett.} \bibinfo{volume}{B 115} (\bibinfo{year}{1982})
  \bibinfo{pages}{495}.
\bibitem[{Bozzo et~al.(1984)}]{UA4_B}
\bibinfo{author}{M.~Bozzo}, et~al. (\bibinfo{collaboration}{UA4
  Collaboration}), \bibinfo{journal}{Phys. Lett.} \bibinfo{volume}{B 147}
  (\bibinfo{year}{1984}) \bibinfo{pages}{385}.
\bibitem[{Arnison et~al.(1983)}]{UA1_B}
\bibinfo{author}{G.~Arnison}, et~al. (\bibinfo{collaboration}{UA1
  Collaboration}), \bibinfo{journal}{Phys. Lett.} \bibinfo{volume}{B 128}
  (\bibinfo{year}{1983}) \bibinfo{pages}{336}.
\bibitem[{Augier et~al.(1993)}]{UA4/2_B}
\bibinfo{author}{C.~Augier}, et~al. (\bibinfo{collaboration}{UA4/2
  Collaboration}), \bibinfo{journal}{Phys. Lett.} \bibinfo{volume}{B 316}
  (\bibinfo{year}{1993}) \bibinfo{pages}{448}.
\bibitem[{Bueltmann et~al.(2004)}]{PP2PP}
\bibinfo{author}{S.~L. Bueltmann}, et~al. (\bibinfo{collaboration}{PP2PP
  Collaboration}), \bibinfo{journal}{Phys. Lett.} \bibinfo{volume}{B 579}
  (\bibinfo{year}{2004}) \bibinfo{pages}{245}.
  \href{http://arxiv.org/abs/nucl-ex/0305012}{\tt arXiv:nucl-ex/0305012}.
\bibitem[{Amos et~al.(1992)}]{E710_B}
\bibinfo{author}{N.~A. Amos}, et~al. (\bibinfo{collaboration}{E710
  Collaboration}), \bibinfo{journal}{Phys. Rev. Lett.} \bibinfo{volume}{68}
  (\bibinfo{year}{1992}) \bibinfo{pages}{2433}.
\bibitem[{Abazov et~al.(2012)}]{D0}
\bibinfo{author}{V.~M. Abazov}, et~al. (\bibinfo{collaboration}{D0
  Collaboration}), \bibinfo{journal}{Phys. Rev.} \bibinfo{volume}{D 86}
  (\bibinfo{year}{2012}) \bibinfo{pages}{012009}.
\bibitem[{Abelev et~al.(2013)}]{ALICE_inel}
\bibinfo{author}{B.~Abelev}, et~al. (\bibinfo{collaboration}{ALICE
  Collaboration}), \bibinfo{journal}{Eur. Phys. J.} \bibinfo{volume}{C 73}
  (\bibinfo{year}{2013}) \bibinfo{pages}{2456}.
  \href{http://arxiv.org/abs/1208.4968}{\tt arXiv:1208.4968}.
\bibitem[{{CMS Collaboration}(2013)}]{CMS_inel}
\bibinfo{author}{{CMS Collaboration}}, \bibinfo{journal}{Phys. Lett.}
  \bibinfo{volume}{B 722} (\bibinfo{year}{2013}) \bibinfo{pages}{5}.
  \href{http://arxiv.org/abs/1210.6718}{\tt arXiv:1210.6718}.
\bibitem[{Engel(1995)}]{PHOJET1}
\bibinfo{author}{R.~Engel}, \bibinfo{journal}{Z. Phys.} \bibinfo{volume}{C 66}
  (\bibinfo{year}{1995}) \bibinfo{pages}{203}.
\bibitem[{Bopp et~al.(1998)Bopp, Engel, and Ranft}]{PHOJET2}
\bibinfo{author}{F.~W. Bopp}, \bibinfo{author}{R.~Engel},
  \bibinfo{author}{J.~Ranft}, \bibinfo{journal}{Conf. Proc. LISHEP98}
  (\bibinfo{year}{1998}) \bibinfo{pages}{729}.
  \href{http://arxiv.org/abs/hep-ph/9803437}{\tt arXiv:hep-ph/9803437}.
\bibitem[{{ATLAS Collaboration}(2012)}]{Rapgap}
\bibinfo{author}{{ATLAS Collaboration}}, \bibinfo{journal}{Eur. Phys. J.}
  \bibinfo{volume}{C 72} (\bibinfo{year}{2012}) \bibinfo{pages}{1926}.
  \href{http://arxiv.org/abs/1201.2808}{\tt arXiv:1201.2808}.

\end{thebibliography}

\onecolumn
\clearpage
\begin{flushleft}
{\Large The ATLAS Collaboration}

\bigskip

G.~Aad$^{\rm 84}$,
B.~Abbott$^{\rm 112}$,
J.~Abdallah$^{\rm 152}$,
S.~Abdel~Khalek$^{\rm 116}$,
O.~Abdinov$^{\rm 11}$,
R.~Aben$^{\rm 106}$,
B.~Abi$^{\rm 113}$,
M.~Abolins$^{\rm 89}$,
O.S.~AbouZeid$^{\rm 159}$,
H.~Abramowicz$^{\rm 154}$
% ...
\bigskip
\\
% ...
$^{11}$ Institute of Physics, Azerbaijan Academy of Sciences, Baku, Azerbaijan\\
% ...
$^{84}$ CPPM, Aix-Marseille Universit{\'e} and CNRS/IN2P3, Marseille, France\\
% ...
$^{89}$ Department of Physics and Astronomy, Michigan State University, East Lansing MI, United States of America\\
% ...
$^{106}$ Nikhef National Institute for Subatomic Physics and University of Amsterdam, Amsterdam, Netherlands\\
% ...
$^{112}$ Homer L. Dodge Department of Physics and Astronomy, University of Oklahoma, Norman OK, United States of America\\
$^{113}$ Department of Physics, Oklahoma State University, Stillwater OK, United States of America\\
% ...
$^{116}$ LAL, Universit{\'e} Paris-Sud and CNRS/IN2P3, Orsay, France\\
% ...
$^{152}$ Institute of Physics, Academia Sinica, Taipei, Taiwan\\
% ...
$^{154}$ Raymond and Beverly Sackler School of Physics and Astronomy, Tel Aviv University, Tel Aviv, Israel\\
% ...
$^{159}$ Department of Physics, University of Toronto, Toronto ON, Canada\\
% ...
\end{flushleft}


\end{document}